\documentclass[fleqn,usenatbib]{mnras}

\usepackage{newtxtext,newtxmath}

\usepackage[T1]{fontenc}
\usepackage{ae,aecompl}

\usepackage{graphicx}	
\usepackage{amsmath}	
\usepackage{amssymb}	
\usepackage{comment}
\usepackage{defs}

\let\oldhref\href
\renewcommand{\href}[2]{\oldhref{#1}{\hbox{#2}}}

\def\hide#1{}

\title[Ghostly Halos]{Ghostly Halos in Dwarf Galaxies: Constraints on the Star Formation Efficiency before Reionization}

\author[Hoyoung Kang and Massimo Ricotti]{
Hoyoung Kang$^{1}$\thanks{E-mail: ghdud14DT@hotmail.com} 
\ and Massimo Ricotti$^{2}$\thanks{E-mail: ricotti@astro.umd.edu} 
\\
$^{1}$Department of Physics, University of Maryland, College Park, MD 20742, USA\\
$^{2}$Department of Astronomy, University of Maryland, College Park, MD 20742, USA
}

\date{Accepted XXX. Received YYY; in original form ZZZ}

\pubyear{2018}

\begin{document}
\label{firstpage}
\pagerange{\pageref{firstpage}--\pageref{lastpage}}
\maketitle

\begin{abstract}
Stellar halos observed around normal galaxies are extended and faint stellar structures formed by debris of tidally disrupted dwarf galaxies accreted over time by the host galaxy. Around dwarf galaxies, these stellar halos may not exist if all the accreted satellites are dark halos without stars. However, if a stellar halo is found in sufficiently small mass dwarfs, the whole stellar halo is composed of tidal debris of fossil galaxies, and we refer to it as ghostly halo. Fossil galaxies are called so because they formed most of their stars before the epoch of reionization, and have been identified as the ultra-faint dwarf galaxies found around the Milky Way and M31. 

In this paper we carry out semi-analytical simulations to characterize the sizes and stellar masses of ghostly stellar halos in dwarf galaxies as a function of their dark matter halo mass. By comparing the models to observations of six isolated dwarf galaxies in the Local Group showing evidence of extended stellar halos, we are able to constrain the star formation efficiency in fossil galaxies. We find that at redshift $z\sim 6$, dark matter halos in the mass range $10^7-10^9$~M$_\odot$ have a mean star formation efficiency $f_* \equiv M_*/M_{dm} \sim 0.05\%-0.1\%$ nearly constant as a function of the dark matter halo mass. There is a tentative indication of a sharp increase of $f_*$ at $M_{dm} \sim 10^6-10^7$~~M$_\odot$, but it is very uncertain and based on only one dwarf galaxy (Leo~T).
\end{abstract}




\section{Introduction}

In cold dark matter (CDM) cosmology, where the term "cold" refers to non-relativistic dark matter, galaxy formation is hierarchical: galaxies grow by accreting smaller mass galaxies. The accretion of these galaxies produces an extended stellar halo, with a stellar mass that depends on the stellar masses of the accreted galaxies and a size that mainly depends on the number of mergers \citep{Eggen1962,Searle1978,White1991,Johnston2008}.
The stellar halo, which extends well beyond the bulge and disk components of a galaxy, consists of stars tidally stripped from accreted dwarf galaxies and is embedded in an even larger halo of dark matter. A stellar halo is observed around the Milky Way, in other spiral galaxies and in galaxy clusters, but not much is known about stellar halos around dwarf galaxies. Here, we define dwarf galaxies to have dark matter halo mass, $M_{\rm dm} < 10^{10}M_{\odot}$, {\it i.e.,} about 100 times smaller than the Milky Way. 

After reionization, at redshift $z \sim 6$, gas in the universe became hot and the thermal pressure overcame the gravitational pull in galaxies with dark matter halo mass below a critical value $M_{\rm dm}^{\rm cr} \sim 10^{8}-10^{9}$ M$_{\odot}$. Therefore the IGM heating associated with reionization prevents fueling with fresh gas of small dwarf galaxies from the epoch of reionization to redshift $z \sim 1-2$, when some gas accretion becomes possible again even in dwarf galaxies with halo masses below the critical value \citep{Ricotti2009}. This process is known as reionization feedback \citep{Babul92,Efstathiou92b,Gnedin00b,Bullock00,OkamotoGT08}.

In dwarf galaxies, most of the stars in the stellar halo have been stripped from tiny dwarfs with total masses $<10^8-10^9$~M$_\odot$. Thus, merging subunits are sufficiently small to be subject to reionization feedback, meaning that most of the stars in the accreted dwarfs, and therefore in the stellar halo, formed before the reionization epoch at $z > 6$. Such dwarf galaxies, dominated by only old stellar populations, are categorized as "true fossils" of the pre-reionization epoch \citep{RicottiG2005,BovillR2009}.
Stellar halos around dwarf galaxies with mass $<10^9-10^{10}$~M$_\odot$, if ever found, are expected to be assembled by multiple mergers with fossil galaxies, and to be a powerful probe of galaxy formation before reionization. These halos will be characterized by an old and metal poor stellar population and an extremely low surface brightness, making them hard to observe. We refer to these dim halos as "ghostly halos" 
\hbox{\citep{BovillR2011a,BovillR2011b}}, being "ethereal" and solely made from dead fossil galaxies.

In this paper we theoretically model the properties of ghostly halos, such as their half-light radii, stellar masses, and stellar mass surface densities. Next, we collect a set of data on isolated dwarf galaxies around the Milky Way that shows evidence of extended old stellar halos. Finally, we compare the observations to the models and show that it is possible to put new constraints on the star formation efficiency and stellar mass in the first galaxies as a function of their dark matter halo mass. This new method is complementary to previous attempts to constrain the properties of the first galaxies using observations of ultra-faint dwarf (UFD) galaxies around the Milky Way and M31, but, of course, should also produce results that are consistent because fossil galaxies and ghostly halos are two manifestations of the same underlaying physical processes.

This paper is organized as follows. In \S~\ref{sec:sim} we describe a simple model for the build up of stellar halos of fossil stars. In \S~\ref{sec:res}, using the model in \S~\ref{sec:sim} and a merger-tree code we present the results of numerical modeling of the assembly of stellar halos as a function of the mass of the galaxy and the assumed star formation efficiency in the first galaxies. In \S~\ref{sec:obs} we review the observational data on the existence and properties of extended stellar halos in nearby isolated dwarf galaxies. In \S~\ref{sec:constr} we compare observations to theoretical modeling in order to constrain the star formation efficiency in the first galaxies. A summary and conclusions are presented in \S~\ref{sec:sum}. Throughout this paper we use Planck \citep{Planck2018} cosmological parameters ($\Omega_m=0.31, \Omega_\Lambda=0.69, h=0.67, \Omega_b=0.049, n_s=0.965, \sigma_8=0.81$).

\section{Simulations}\label{sec:sim}

In order to estimate the growth of the stellar halo produced by repeated mergers, we adopt the formalism in \cite{BK2005,BK2006} used to describe the build up of bulges as a result of minor and major merger events. The final effective radius of the stellar halo, $R_f$ (which we can define as the radius at which half of the total luminosity is emitted) produced by a binary merger of two galaxies can be related to their initial radii $R_1, R_2$ by the energy conservation relationship:
\begin{align}
f_{f}\frac{M_{f}^2}{R_{f}}&=f_{1}\frac{M_{1}^2}{R_{1}}+f_{2}\frac{M_{2}^2}{R_{2}}+(f_{orb}+f_{t})\frac{M_{1}M_{2}}{R_{1}+R_{2}},
\end{align}
where $M_{1}$ and $M_{2}$ are initial masses of the stellar spheroids, $M_f=M_1+M_2$ is the final stellar halo mass, $f_{orb}$ contains information about the orbital energy of the merger event, and $f_{t}$ describes the energy transfer between the stellar and the dark matter components. The parameters of order of unity $f_1, f_2$ and $f_f$, depend on the detail of the dark matter and stellar structure of the initial galaxies and the final galaxy, respectively. These parameters can only be obtained using numerical simulations as they encode both the gravitational potential energy and the internal kinetic energy of the galaxy.
However, if we assume perfect homology ({\it i.e.}, the profiles of galaxies are the same up to scaling constants) and a parabolic orbit with no energy transfer between the dark matter and the stellar component, then $f_{f}=f_{1}=f_{2}$ and $f_{orb}=f_{t}=0$, respectively \citep{BK2005, Novak}. With these assumptions we obtain the following simplified equation:
\begin{align}
\frac{M_{f}^2}{R_{f}}=\frac{M_{1}^2}{R_{1}}+\frac{M_{2}^2}{R_{2}}.
\end{align}
If we further assume a constant radius, $R_0$, for each initial halo mass and we consider the merging of $N$ galaxies, $M_{i}$ for $1\le i\le N$, then we can solve for $R_{f}$:
\begin{align}
\frac{R_{f}}{R_{0}}={M_{f}^2}/{\sum\limits_{i=1}^N M_{i}^2}.\
\label{eq:simple}
\end{align}
Here, $M_{f}\equiv \sum_{i=1}^N M_{i}$ is the final stellar mass of the halo and $M_{i}$ is the stellar mass of the merging galaxy $i$. We can better understand the physical meaning of this equation considering two extreme cases: i) If all the $N$ merging galaxies have equal mass, then $R_{f}=N R_0$; ii) If only one of the galaxies dominates the summation, then $R_{f}=R_0$. Therefore, numerous minor mergers are more effective than a single major merger in increasing the size of the stellar halo. Here, we have used Equation~(\ref{eq:simple}) only for illustration purposes, because the assumption that the effective radii of galaxies do not depend on their masses is clearly unrealistic.

\begin{figure}
\centering
\includegraphics[width=0.4\textwidth]{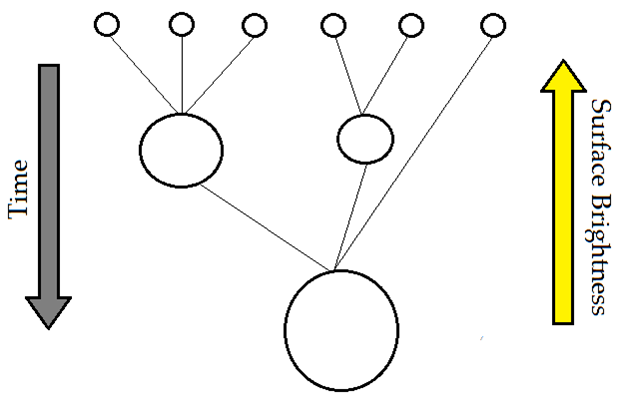}
\caption{A schematic of a halo merger tree as in the Monte-Carlo merger tree code used here. Size of the nodes represent the size of the galaxies. As time increases, redshift decreases. Since we assume there are no newly formed stars in small mass halos after reionization, surface brightness decreases as galaxies merge.}\label{fig:merg}
\end{figure}

Hereafter, we will assume that the effective stellar radius of each merging galaxy depends on its dark matter halo mass as follows:
\begin{align}
R_{i} = R_{0}\left(\frac{M_{dm,i}}{M_{dm,0}}\right)^{\alpha},
\label{eq:rh}
\end{align}
where we adopt an effective radius $R_0=0.13$~kpc for a halo mass $M_{dm,0}=10^{7}$~M$_\odot$, and $\alpha=1/3$. This assumption is based on observations of late type galaxies \citep{Kravtsov2013} and can be interpreted in the context of the theory of angular momentum conservation and tidal torques. Moreover, the relationship states that the half-light radius of the stellar component ($r_{ h}$) is a constant fraction of the virial radius of the dark matter halo ($R_{200}$). For early and late type galaxies with halo masses $10^9~M_\odot<M_{dm}<10^{14}~M_\odot$, the half-light radii follow the linear relationship $r_h=0.015 R_{200}$, but the scatter around the mean is a factor of $\pm 300\%$. Here we adopt the linear relationship $r_h=0.03R_{200}$, a factor of two higher than the mean, because it is a better fit when considering only late type dwarf galaxies, which are the focus of this paper. In addition we will relate the stellar mass of a galaxy to the total dark matter halo mass,
\begin{align}
M_i=f_{*,i}M_{dm,i}\label{eq:ms},\ 
\end{align}
where the star formation efficiency, $f_*$, is parameterized as a power-law:
\begin{align}
f_{*,i}=\epsilon_0\left(\frac{M_{dm,i}}{M_{dm,0}}\right)^{\beta}\label{eq:fs},
\end{align}
with slope $\beta$, which is a free parameter in the model. The free parameter $\epsilon_0$ ({\it i.e.}, the star formation efficiency for halo masses $M_{dm,0}$) is a normalization factor which trivially affects the results, as shown below.
Therefore, the generalization of Equation~(\ref{eq:simple}) is
\begin{align}
\frac{R^{\rm halo}}{R_0}&=\frac{\left(\sum\limits_{i=1}^N \mu_{i}\right)^2}{\sum\limits_{i=1}^N \mu_{i}^{2-\frac{\alpha}{\beta+1}}}=\frac{\left(\sum\limits_{i=1}^N M_{dm,i}^{\beta+1}\right)^2}{M_{dm,0}^{\alpha}\sum\limits_{i=1}^N M_{dm,i}^{2(\beta+1)-\alpha}},
\label{eq:rf}
\end{align}
where $\mu_i \equiv M_{i}/M_{0}$ and $M_{0}=\epsilon_0 M_{dm,0}$.
Note that $R^{\rm halo}$ depends only on the merger tree history of luminous halos, $\beta$ (a free parameter) and $\alpha$ (which we set to $1/3$), but it is independent of the assumed free-parameter $\epsilon_0$ and of the choice of the pivot dark matter mass $M_{dm,0}$, because $R_0/M_{dm,0}^\alpha=R_i/M_{dm,i}^\alpha$.
Finally, a very simple expression can be derived for the ratio of the halo effective radius to the galaxy effective radius, $R^{\rm gal}=R_0(M_{dm, gal}/M_{dm,0})^\alpha$:
\begin{align}
\frac{R^{\rm halo}}{R^{\rm gal}}&=\frac{\left(\sum\limits_{i=1}^N M_{dm,i}^{\beta+1}\right)^2}{M_{dm,gal}^{\alpha}\sum\limits_{i=1}^N M_{dm,i}^{2(\beta+1)-\alpha}},
\label{eq:rf1}
\end{align}
showing clearly that this ratio (which, as we will see later, can be obtained from observations of isolated dwarf galaxies) can be used to constrain $\beta$ and therefore the star formation efficiency at the epoch of reionization. In order to estimate $R^{\rm halo}$ in Equation~(\ref{eq:rf}) we need to know the merger history of a galaxy (see Fig.~\ref{fig:merg} for a sketch). We use a Monte-Carlo merger tree code \citep{Parkinson} based on the extended Press-Schechter formalism to produce several realizations of the merger history of galaxies residing in a dark matter halo of mass $M_{dm,gal}$.
For each halo mass at $z=0$ and set of free parameters, we report the mean and the standard deviation of each observable based on $60$ random merger-tree realizations to account for the different ways the final galaxy may be assembled.
Since we are interested in determining the mass and radius of the stellar halo in isolated dwarf galaxies at $z=0$, we assume that the merging building blocks have dark matter masses sufficiently small to be affected by reionization feedback. Our  assumption is justified as long as the merging galaxies have DM masses $<M_{J}$, where $M_{J} \sim 10^9 M_\odot /(1+z_{\rm merge})$ is roughly the Jeans mass of the IGM after reionization, assuming an IGM temperature $\sim 10^4$~K. Since halos with mass $<M_{J}$ can only form stars before reionization, we populate dark matter halos with stars roughly at the epoch of reionization $z \sim z_{\rm rei}$ according to the star formation law in Equation~(\ref{eq:fs}). It is important to emphasize that our free parameters for the star formation efficiency ($\epsilon_0$ and $\beta$) are not the values at present time but how they were at the epoch of reionization, or more precisely at the epoch when reionization feedback shuts down star formation in dwarf galaxies\footnote{The exact redshift when star formation is suppressed in small dwarf galaxies may differ by a few galactic dynamical times with respect to the local value of $z_{\rm rei}$ because, in small mass halos internal, photoionization may sterilize a galaxy earlier or star formation may continue for a short time after $z_{\rm rei}$ until full consumption of existing molecular gas in the interstellar medium.}. Since these halos cannot form stars after reionization, we simply follow their merger history with the host halo to $z=0$. So, the stellar mass of the halo and its radius are simply set by the volume density distribution function of stellar masses of the building blocks at $z=z_{\rm rei}$.
Clearly, this model fails when the building blocks of the stellar halo have mass $>M_{\rm J}$, which roughly corresponds to dwarf galaxies with present-day dark matter halo masses $> 10^{10}$~M$_\odot$.

In our simulations, redshift values for the onset of reionization feedback are chosen at $z_{\rm cut} = 6$ and $z_{\rm cut} = 5$ because reionization of the universe was completed around $z_{\rm rei}=6.2$ at $t \sim 900$~Myr, and we investigate the effect of negligible delay ($50$~Myr) and  longer delay ($300$~Myr) of this feedback with respect to $z_{\rm rei}$. The simulations span a range of present-day dark matter halo masses from $10^{8}$~M$_{\odot}$ to $10^{11}$~M$_{\odot}$ to cover a wide range of dwarf galaxy masses. Additionally, we explore a range for the free parameter $\beta$ from -0.5 to 2 in order to capture possible star formation efficiencies. Another important parameter in the simulation is the mass resolution of the merging halos, which sets a limit to the smallest (luminous) dwarf galaxy that can merge. We explore two cases for the minimum halo mass of the merging building blocks: $M_{\rm res}=10^6$~M$_{\odot}$ and $M_{\rm res}=5 \times 10^6$~M$_{\odot}$.

\section{Modeling Results}\label{sec:res}
\begin{figure*}
\centering
\includegraphics[width=0.32\textwidth]{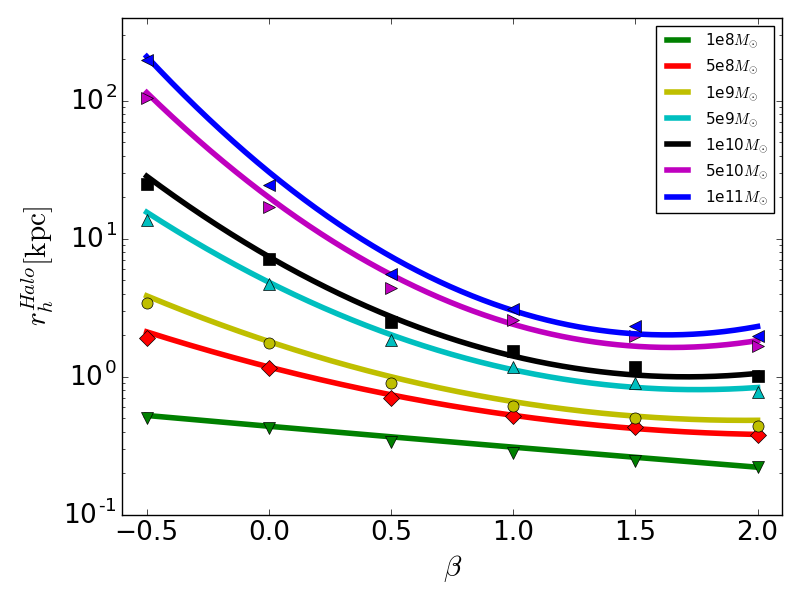}
\includegraphics[width=0.32\textwidth]{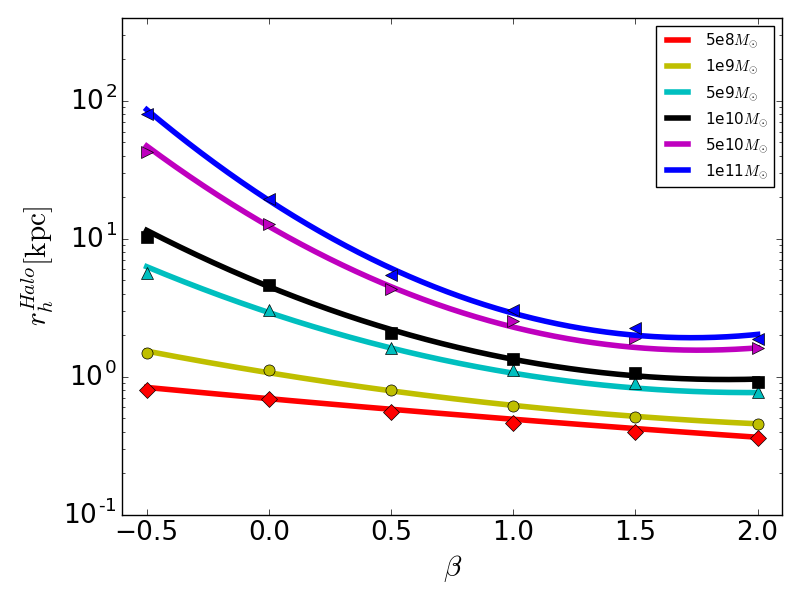}
\includegraphics[width=0.32\textwidth]{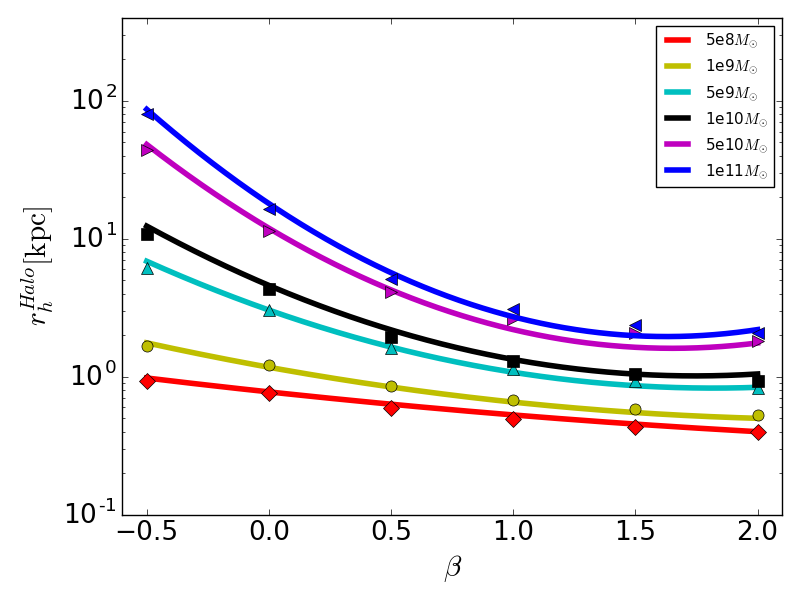}\\
\includegraphics[width=0.32\textwidth]{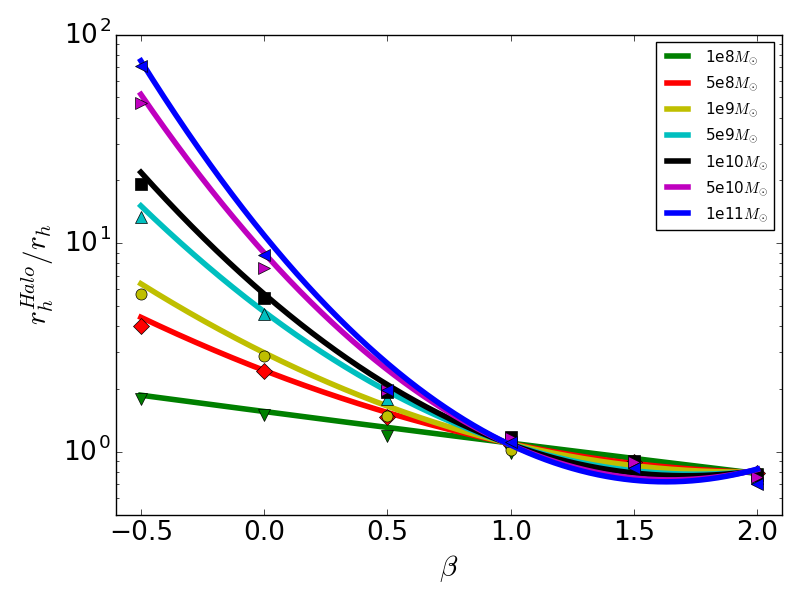}
\includegraphics[width=0.32\textwidth]{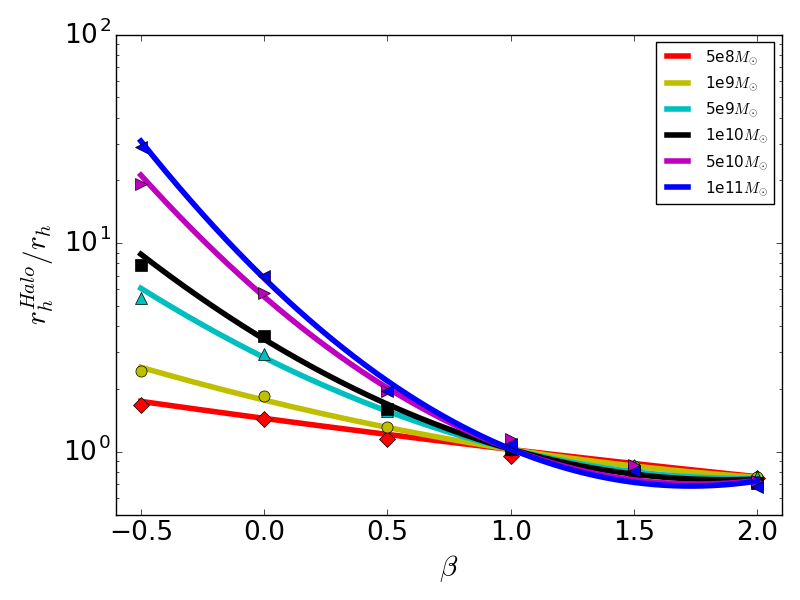}
\includegraphics[width=0.32\textwidth]{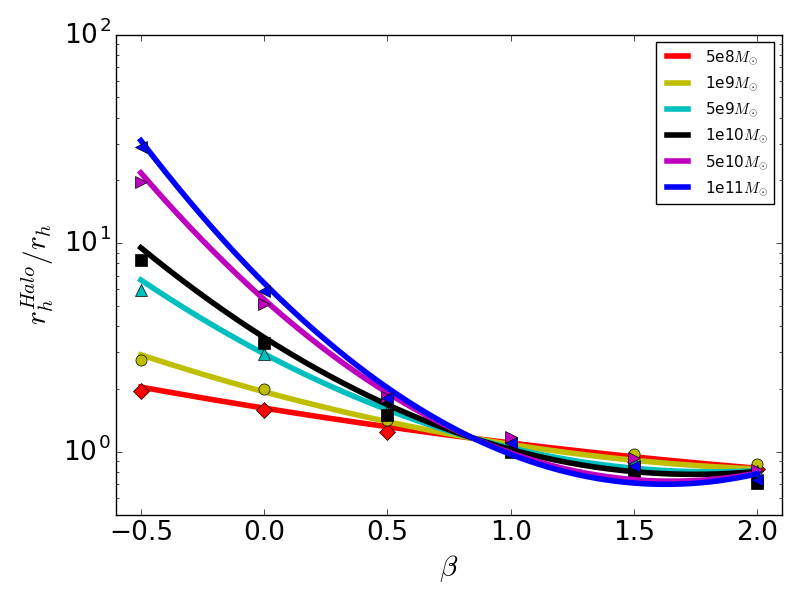}
\caption{{\it (Top panels.)} Average final effective radii of stellar halos, $r_{\rm h}^{\rm halo}$ as a function of $\beta$ and $M_{\rm dm}$, from 60 random realization of merger histories. In the left panel the sub-halos stellar properties are set at redshift $z_{\rm cut} =6$ and the minimum dark matter halo mass hosting a galaxy is $M_{\rm res}=10^6 M_{\odot}$. In the middle panel $z_{\rm cut}=6$ and $M_{\rm res}=5 \times 10^6 M_{\odot}$. In the right panel $z_{\rm cut}=5$ and $M_{\rm res}=5 \times 10^6 M_{\odot}$. In the legends we report the dark matter halo masses of galaxies at the present day. The data points from the simulations are fit as a function of dark matter halo mass and $\beta$. {\it (Bottom panels.)} Same as the top panels but showing the ratio $r_{\rm h}^{\rm halo}/r_{\rm h}^{\rm gal}$.}\label{fig:radius}
\end{figure*}

\begin{figure*}
\centering
\includegraphics[width=0.32\textwidth]{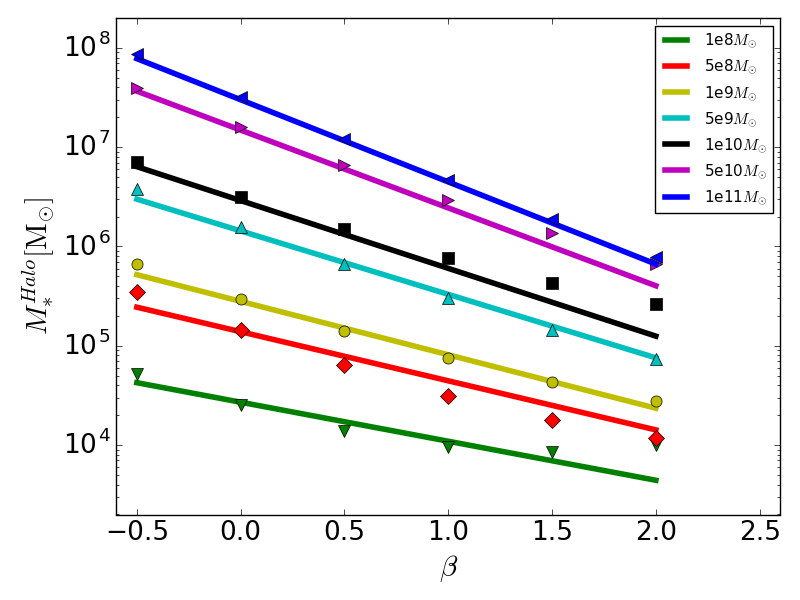}
\includegraphics[width=0.32\textwidth]{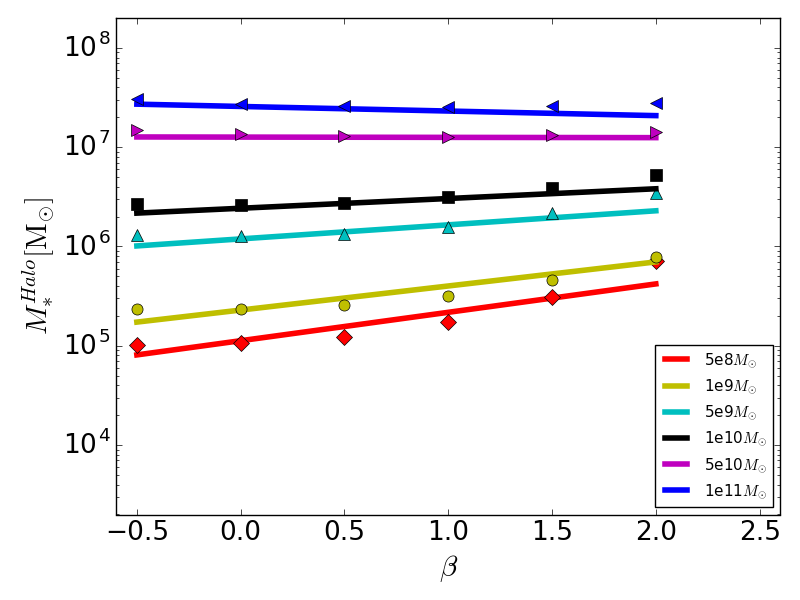}
\includegraphics[width=0.32\textwidth]{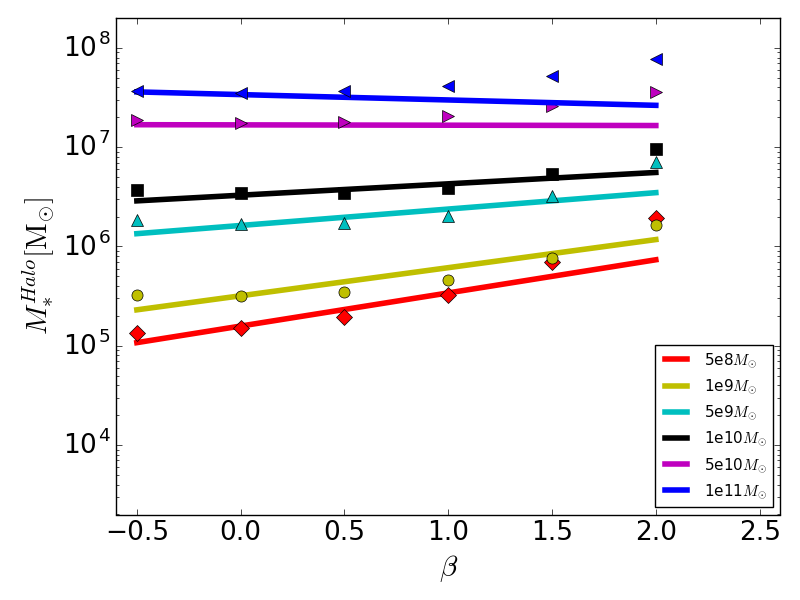}\\
\includegraphics[width=0.32\textwidth]{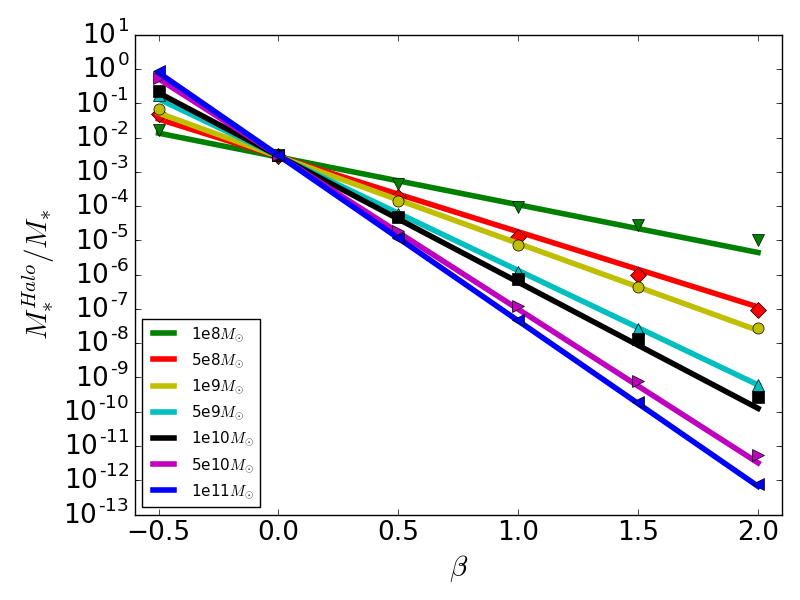}
\includegraphics[width=0.32\textwidth]{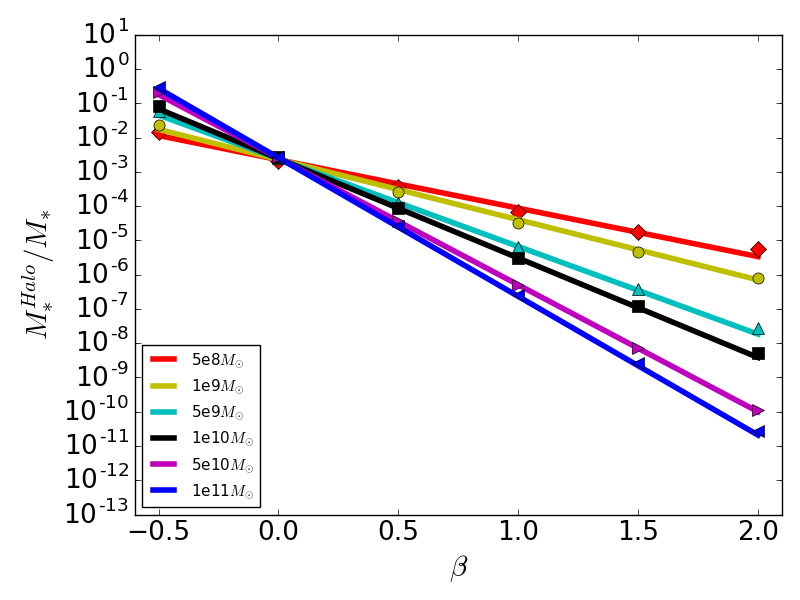}
\includegraphics[width=0.32\textwidth]{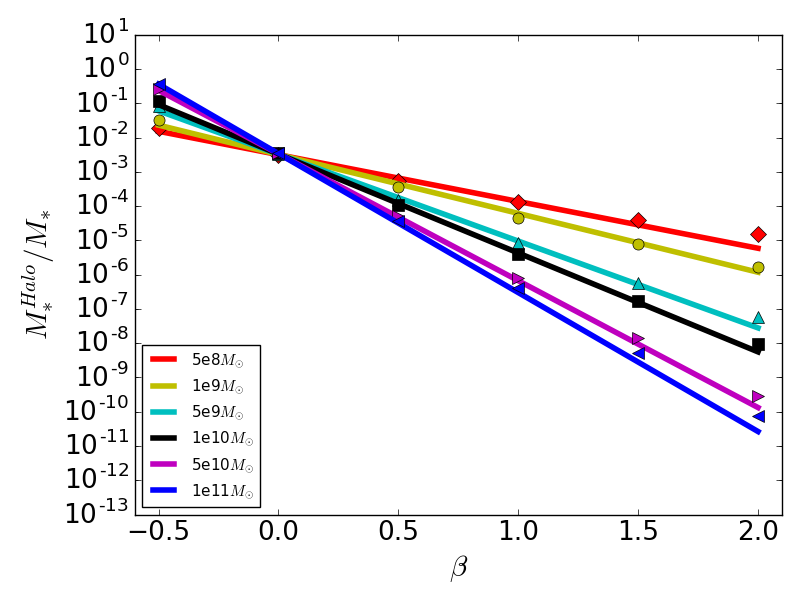}
\caption{{\it (Top panels.)} Same as Fig.\ref{fig:radius} but showing the average mass of stellar halos, $M_*^{\rm halo}$, as a function of $\beta$ and $M_{\rm dm}$.  {\it (Bottom panels.)} Same as the top panels but showing the ratio $M_*^{\rm halo}/M_*^{\rm gal}$.}\label{fig:mass}
\end{figure*}
\begin{figure*}
\centering
\includegraphics[width=0.32\textwidth]{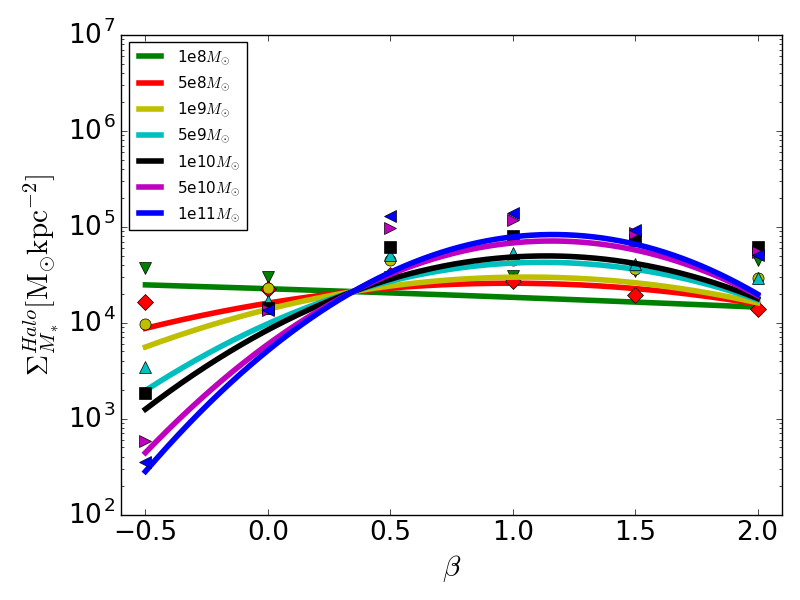}
\includegraphics[width=0.32\textwidth]{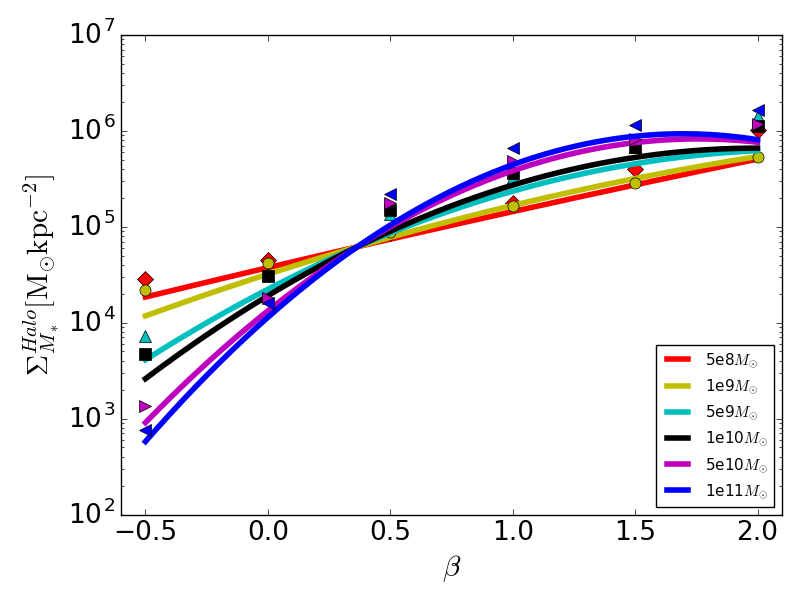}
\includegraphics[width=0.32\textwidth]{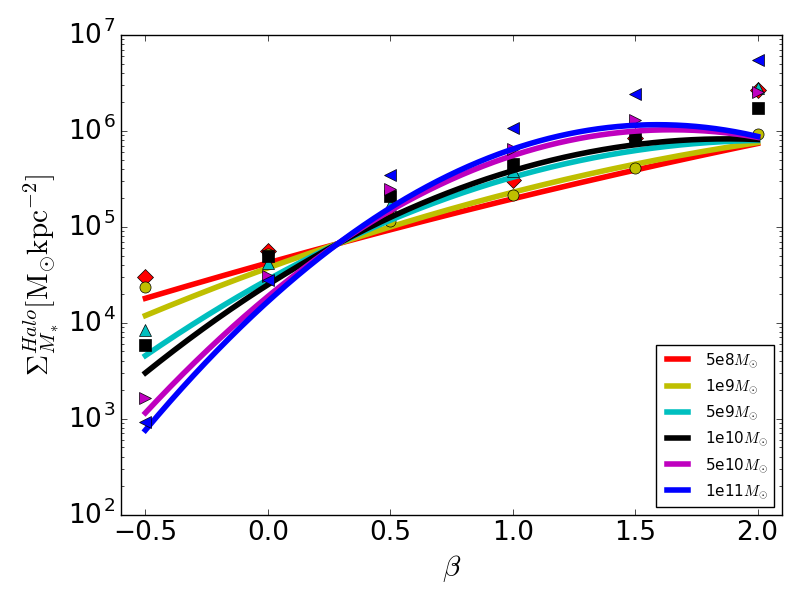}\\
\includegraphics[width=0.32\textwidth]{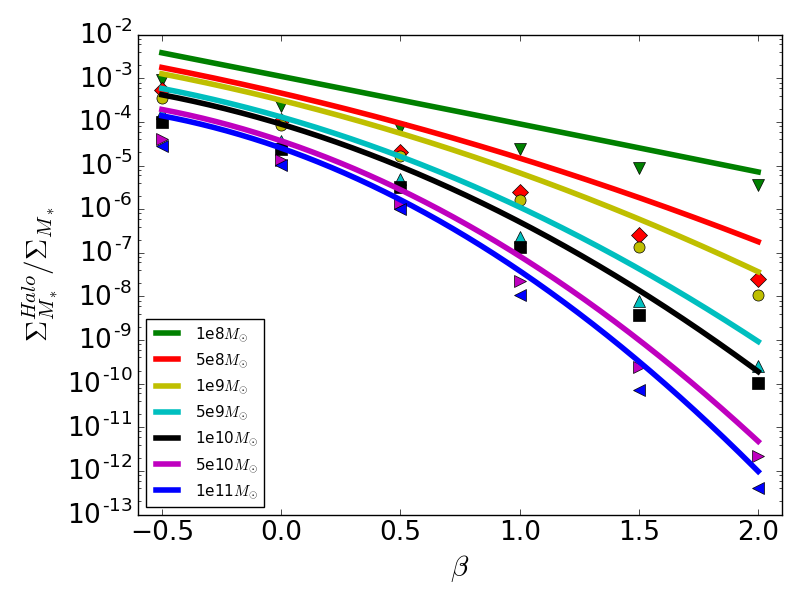}
\includegraphics[width=0.32\textwidth]{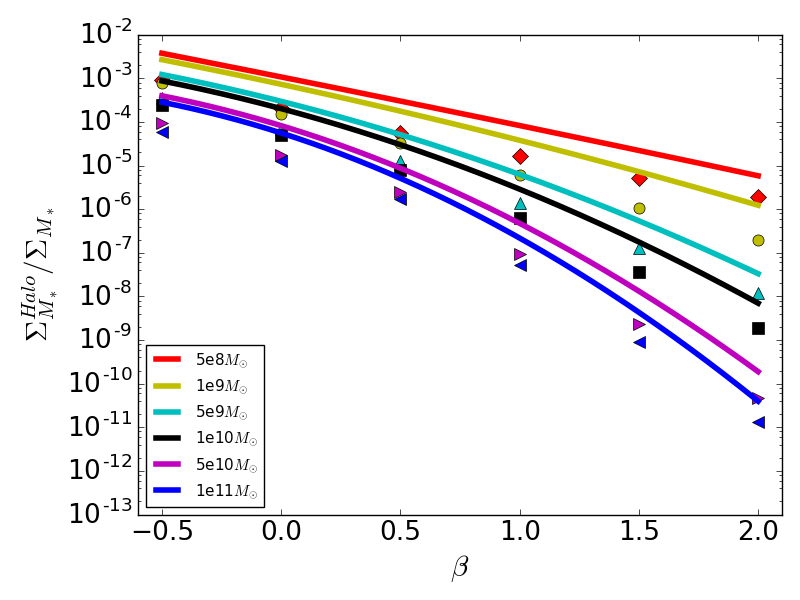}
\includegraphics[width=0.32\textwidth]{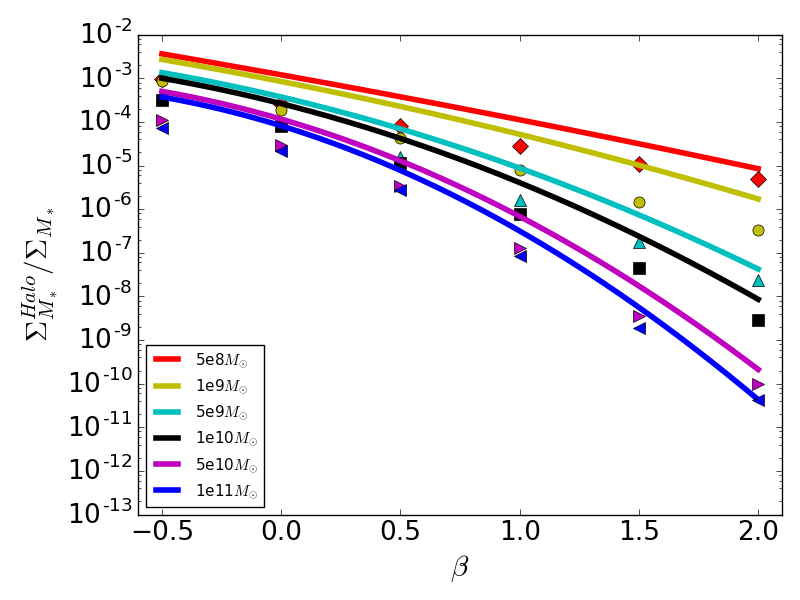}
\caption{{\it (Top panels.)} Same as Fig.\ref{fig:radius} but showing the average surface density of stellar halos, $\Sigma_{*}^{\rm halo}$, as a function of $\beta$ and $M_{\rm dm}$. {\it (Bottom panels.)} Same as the top panels but showing the ratio $\Sigma_{*}^{\rm halo}/\Sigma_{*}^{\rm gal}$.}\label{fig:sigma}
\end{figure*}

We carry out simulations to investigate how the free parameters in our model impact the final effective radii of stellar halos, their stellar masses and stellar mass surface densities. For each set of parameters we report the average and standard deviation from 60 random merger tree realizations. We assume Gaussian distribution for estimating the mean and standard deviations, however we find that this is a good approximation only for $\beta \sim 0$, while for $\beta \sim 1-2$ the distribution is skewed. We also notice that the scatter of the results for a fixed $\beta$ becomes smaller lowering the value of $z_{\rm cut}$. This is most likely due to the fact that we expect more ways galaxies can merge starting at higher redshift.

\subsection{Effective Radius, Mass and Surface Density of Ghostly Stellar halos}

The results from the merger-tree simulations are shown in Figures~\ref{fig:radius}-\ref{fig:sigma}. The points show the mean values from 60 random merger history realizations, while the lines show the best fits assuming polynomial functions (see Appendix~\ref{app:fit}). Figure~\ref{fig:radius} shows the mean half-light radii of stellar halos, $r^{\rm halo}_{ h}$ (top panels), and $r^{\rm halo}_{ h}/r^{\rm gal}_{h}$ (bottom panels) as a function of $\beta$ for different dark matter halo masses at $z=0$ (as shown in the legend). The figures on the left, center and right refer to different assumptions on the parameters $z_{\rm cut}$ and $M_{\rm res}$, and are shown to illustrate the sensitivity of the results to the assumed parameters. The fiducial runs are shown on the left panels and have $z_{\rm cut}=6$ and $M_{\rm res}=10^6$~M$_\odot$. The central panels show the effect of increasing $M_{\rm res}$ to $5 \times 10^6$~M$_\odot$ keeping $z_{\rm cut}=6$, and the right panels show the effect of decreasing $z_{\rm cut}$ to $5$, keeping $M_{\rm res}=5 \times 10^6$~M$_\odot$. The mean values of $r^{\rm halo}_{h}$ from the simulations follow a well defined functional form, however the scatter due to different realization of the merger history is typically about $\pm 30$\% for $\beta \simeq 0$, $\pm 20$\% for $\beta \simeq -0.5$ and $\pm 40$\% for $\beta \simeq 0.5$ and halo masses $>10^9$~M$_\odot$ (see Table~\ref{tab:scatter1} in Appendix~\ref{app:fit}). As expected from Equation~(\ref{eq:rf}) (and the simplified Eq.~(\ref{eq:simple})), $r_h^{\rm halo}$ increases with decreasing $\beta$ because the stellar halo is assembled by many merging galaxies of similar stellar mass, while for large $\beta$ the most massive mergers dominate the stellar budget in the halo. The dependence on the halo mass can be understood similarly because the number of merging galaxies scales with the halo mass, given our assumption of a minimum mass cutoff on the halo mass, $M_{\rm res}$, representing the transition to halos that form single Pop~III stars rather than primordial dwarf galaxies. The value of $r_{h}^{\rm halo}$ is linearly proportional to the assumed normalization $R_0$ for the half light radius of galaxies. Therefore the uncertainty on the normalization of $r_{h}^{\rm halo}$ can be large given the intrinsic scatter around the mean relationship in Eq.~(\ref{eq:rh}). However, the ratio $r^{\rm halo}_{h}/r^{\rm gal}_{h}$ is independent of $R_0$ and is a more robust diagnostic of $\beta$. The main uncertainty is in how precisely the dark matter halo mass of the observed dwarf can be estimated. The scatter of $r_{h}^{\rm halo}$ between galaxies with the same dark matter halo mass is largely due to different realizations of their merger histories, so it would be useful to have a rather large sample of observed dwarf galaxies to compare to the mean values from the model.

Figure~\ref{fig:mass} is the same as Fig.~\ref{fig:radius} but for the mass of the stellar halo, $M_*^{\rm halo}(\epsilon_0/0.001)$, and the ratio $M_*^{\rm halo}/M_*^{\rm gal}$. For this figure we have assumed for the normalization and pivot point in Equation~(\ref{eq:fs}) the values $\epsilon_0=0.001$ and $M_{dm,0}=10^7$~M$_\odot$. With this choice for the normalization, the dependence of $M_*^{\rm halo}$ on $\beta$ is minimized. This is expected because the halo masses contributing to the stellar halos are in the range $10^6-10^8$~M$_\odot$. It is interesting to note that $M_*^{\rm halo}/M_*^{\rm gal}$ is a quantity rather independent of the dark matter halo mass in which a galaxy resides for the case $\beta \sim 0$.

Figure~\ref{fig:sigma} is the same as Fig.~\ref{fig:radius} but for the surface mass density $\Sigma_*^{\rm halo}(\epsilon_0/0.001)$ 
and the ratio $\Sigma_*^{\rm halo}/\Sigma_*^{\rm gal}$.
Here we define the surface brightness and stellar mass surface density within the half-mass radius:
\begin{align}
\Sigma_* =\frac{L}{2\pi R_{h}^2},~~~~~~
\Sigma_{*}^{\rm halo} =\frac{M_{*}^{\rm halo}}{2\pi R_{h}^2}\label{eq:ss},
\end{align}
respectively. The stellar mass surface density of ghostly halos is rather constant as a function of the halo mass for $0<\beta<0.5$ for our fiducial model parameters (left panels). Therefore $\Sigma_*^{\rm halo}$ is a good observable to estimate $\epsilon_0$.

The dependence of the main observables of ghost halos on the minimum mass of luminous building blocks, $M_{\rm res}$, is quite significant when changing $M_{\rm res}$ from $10^6$ to $5\times 10^6$. While, the results are less sensitive to changing $z_{\rm cut}$ from 6 to 5.

\section{Compiling Observational Data}\label{sec:obs}

\begin{table*}
\begin{center}
    \begin{tabular}{lllllllcll}
    \hline
    Galaxy & Dist. & m-M & m$_V$ & M$_V$ & [Fe/H] & $r_{h}$ & Refs.$^{(a)}$ & Comments    \\ 
    & [kpc] & & & & & [kpc] & & \\ \hline
    
    Leo T & 420 & 23.1 & 16 & -7.1 & -1.6 & 0.170 (0.120)& (1) & MW sub-group; faint as UFD but recent SF.\\
    Leo A & 800 & 24.5 & 12.4 & -12.1& -1.4 & 0.354 (0.499)& (2) (3) & Local Group, isolated; recent SF.\\
    WLM & 985 &25.0 & 10.9& -14.1 & -1.7 & 1.66 (2.111)& (4) (5)& Local Group, isolated; has one bright GC\\
     IC1613 & 758 & 24.4 & 9.5 & -14.9 & -1.26 & 1.502 (1.496))& (6) (7)& M31 sub-group; recent SF; dust poor.\\
     IC10 & 750 &24.4 & 9.4& -15 & -1.3 & 0.612 (0.612) & (8) & M31 sub-group; starburst; huge HI gas envelope. \\
    NGC 6822 & 470 & 23.4 & 8.2 & -15.2 & -1 & 0.478 (0.354)& (9) & MW sub-group; recent SF; metal rich stars in halo.\\ 
    \hline
    \end{tabular}
    \caption{List of dwarf galaxies showing evidence of an extended stellar halo. (a) References: (1) \citet{2007LeoT}; (2) \citet{2004LeoA}; (3) \citet{McConnachie2012}; (4) \citet{2012WLM}; (5)\citet{vandenBergh1994}; (6) \citet{2015IC1613}; (7) \citet{Battinelli2009}; (8) \citet{2015IC10}; (9) \citet{2006NGC6822}. Half-light radii, $r_{h}$, in parenthesis are from \citet{McConnachie2012}, otherwise all the data in the table columns (distance, distance modulus $m-M$, absolute visual magnitude $M_V$, and metallicity [Fe/H]) are from the references cited above. $M_V$ for Leo~A, IC~10 and NGC 6822 are from \citet{McConnachie2012}.}\label{tab:gal1}
    \end{center}    
\end{table*}
\begin{table*}
\begin{center}
    \begin{tabular}{llllllllll}
    \hline
    Galaxy & \multicolumn{3}{c}{Exp. Galaxy Parameters} & \multicolumn{3}{c}{Exp. halo Parameters} & $M_{\rm conv}$ & $\tilde{\chi}^2({\rm halo})$ \\ 
    & $\Sigma_0$~[kpc$^{-2}$] & $r_{\rm exp}$~[kpc] & RG stars & $\Sigma_0$~[kpc$^{-2}$] & $r_{\rm exp}$~[kpc] & RG stars & [$M_{\odot}$] & \\ 
\hline
\\    
  Leo T & $5967\pm 1955$ & $0.05\pm 0.01$ & 112 & $455\pm 237$ & $0.18\pm 0.02$ & 92 & $286.0$ &  3.3\\
  
    Leo A & $10009\pm 1825$ & $0.19\pm0.01$ & 2214 & $337\pm 98$ & $0.52\pm 0.05$ & 574 & $1368.0$ & 1.4 \\
    
      WLM & $2640\pm 683$ & $0.61\pm 0.05$ & 6209 & $55\pm 35$ & $2.7\pm 0.9$ & 2612 & $1368.6$ &  1.4\\
      
       IC1613 & $433\pm 27$ & $0.69\pm0.03$ & 1299 & $9\pm 6$ & $2.9\pm 1$ & 477 & $9157.1$ & 9.1 \\
       
       IC10 & $889\pm 433$ & $0.68 \pm 0.01$ & 25791 & $274\pm 100$ & $1.7\pm0.3$ & 5344 & $1798.2$ & 1.08 \\
       
      NGC6822 & $29459\pm 11063$ & $0.52\pm0.05$ & 51189 & $8 2387\pm 743$ & $1.4\pm0.1$ & 28624 & $339.7$ &  1.15\\  

\end{tabular}
    \caption{Fitting parameters in Fig.~\ref{fig:profs} for the exponential disk and exponential halo profiles. $M_{\rm conv}$ is the parameter derived in Appendix~\ref{app:cmd} to convert from number of RG stars to stellar mass at zero age. }\label{tab:gal2}
    \end{center}    
\end{table*}
We have conducted a literature review on nearby dwarf galaxies located in the Local Group but outside the virial radii of the Milky Way and Andromeda, looking for references in which provided the radial density profile of stars to large distances and showing robust or even tentative evidence of the existence of an extended stellar halo. The selection of dwarfs outside the sphere of influence of the Milky Way and M31 is motivated by the expectation that an extended stellar halo would not survive for long once the dwarf is within the virial radius of a larger host galaxy. The stellar halos are so faint that in nearby dwarf galaxies are typically found by star counts. Usually only the red giants are sufficiently bright to be detected in deep color-magnitude diagrams (CMD).

\begin{figure*}
\centering
\includegraphics[width=0.48\textwidth]{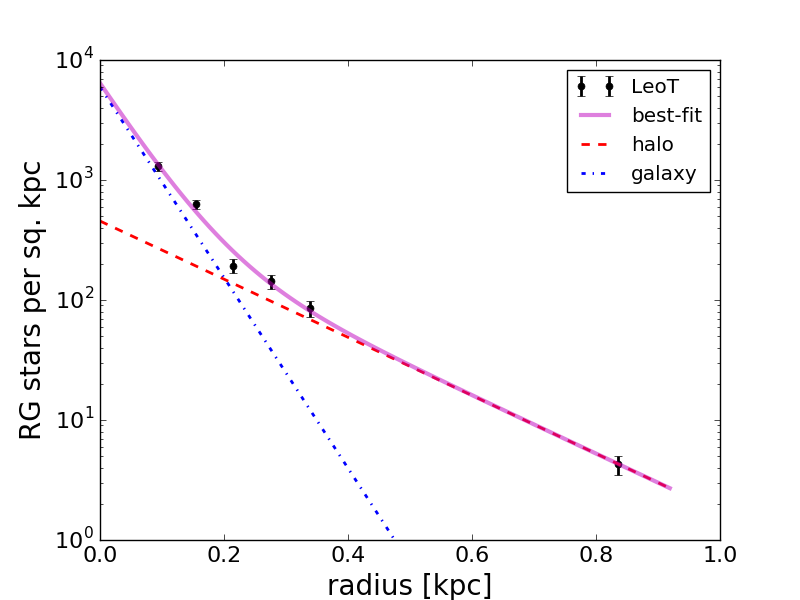}
\includegraphics[width=0.48\textwidth]{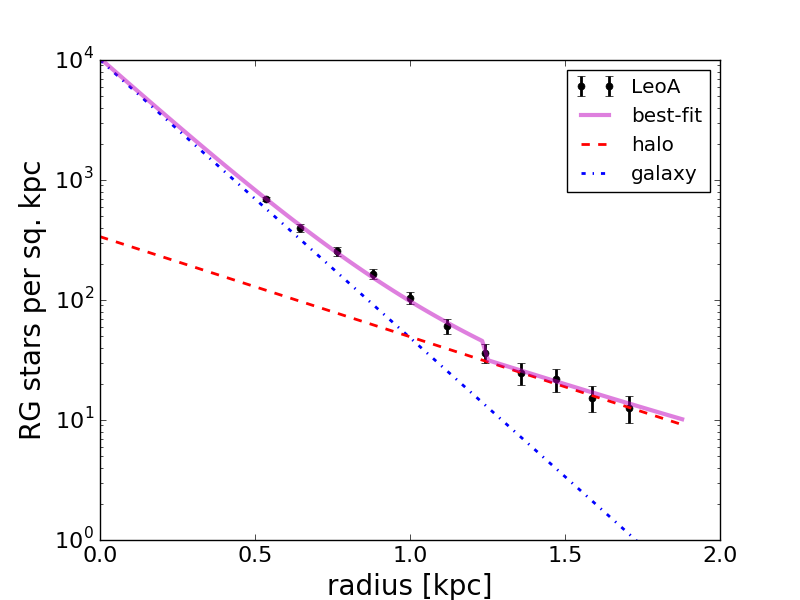}\\
\includegraphics[width=0.48\textwidth]{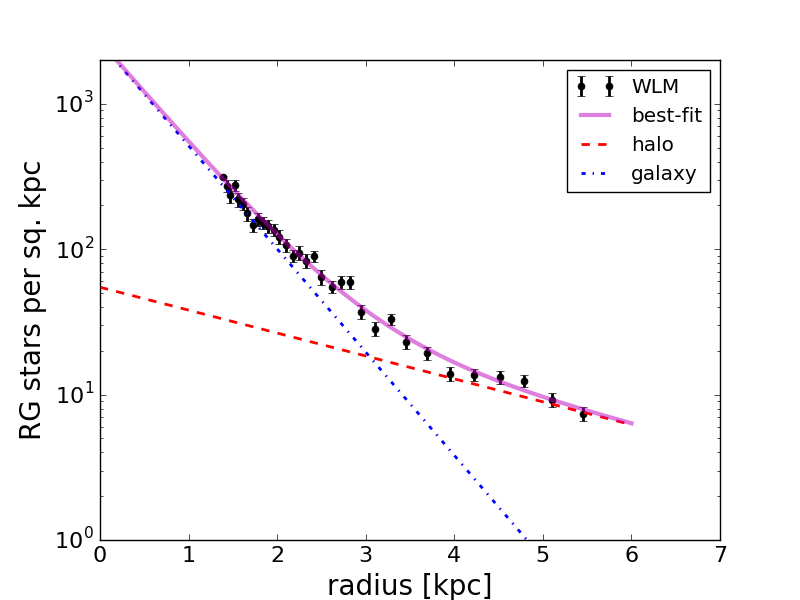}
\includegraphics[width=0.48\textwidth]{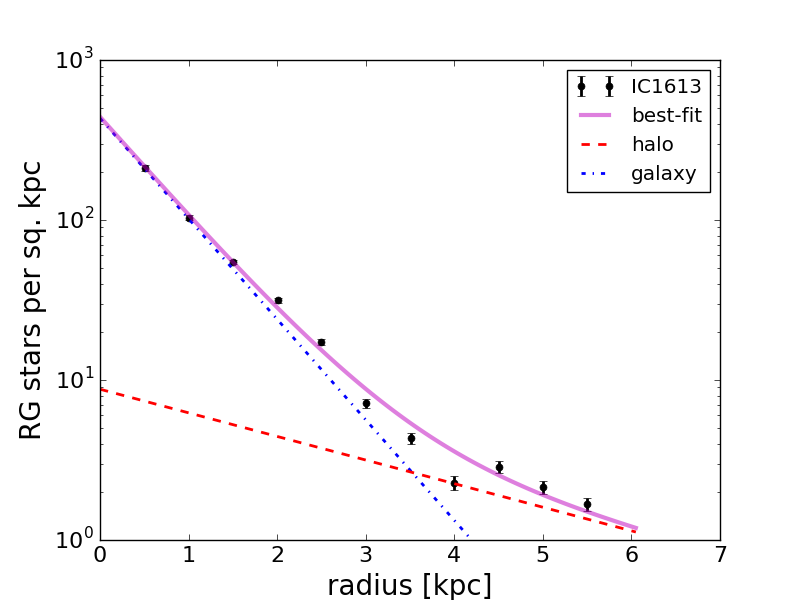}\\
\includegraphics[width=0.48\textwidth]{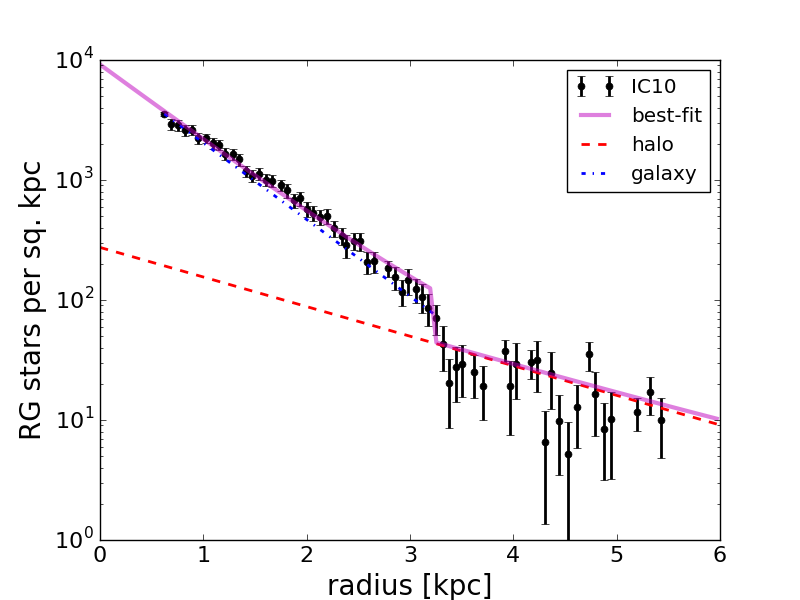}
\includegraphics[width=0.48\textwidth]{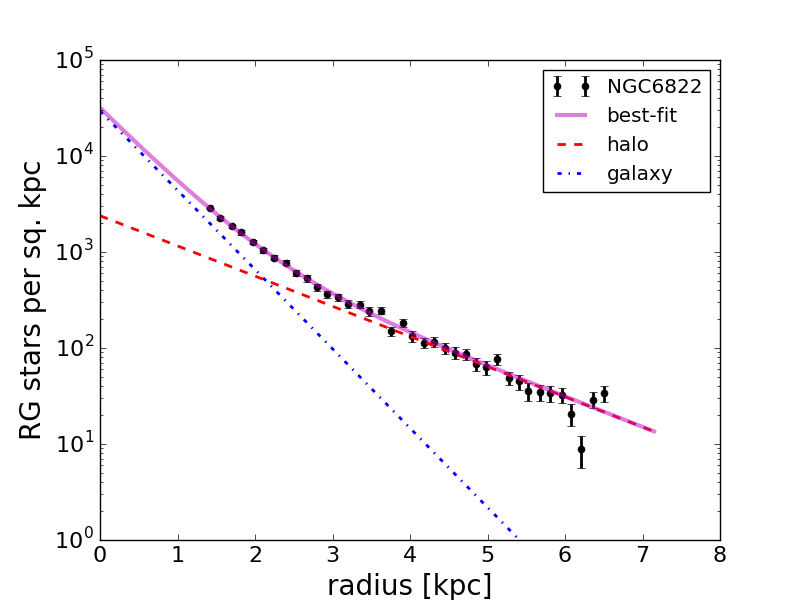}
\caption{Radial density profiles (number of red-giant stars per kpc$^2$) of LeoT \citep{2007LeoT}, LeoA \citep{2004LeoA}, WLM \citep{2012WLM}, IC1613 \citep{2015IC1613}, IC10 \citep{2015IC10}, and NGC6822 \citep{2006NGC6822}. The points with errorbars is the data from the references listed above. We fit the data as the sum of two stellar components with exponential density profiles: the stars in the disk galaxy, and the stars in a more extended stellar halo. The best fit line, which is the sum of the two exponential profiles, is shown as a solid line. The dashed line is the stellar halo profile and the dot-dashed line is the galaxy profile.}\label{fig:profs}
\end{figure*}

We found data for six nearby dwarf galaxies, including Leo~T, Leo~A, WLM, IC1613, IC10 and NGC6822 (see Table~\ref{tab:gal1}). In the literature the data was fit with two exponential profiles, one for the galaxy and one for the extended halo. In some case the fit was given in terms of a single Sersic profile, but here we only used the data points and re-fitted with two exponential profiles, as shown in Figure~\ref{fig:profs}. The fitting is done using a maximum likelihood estimator between the data points (including the uncertainty on the y-axis) and the model being the sum of two exponential profiles. For IC10, because of the break in the profile at about 3~kpc, we fit the exponential profiles independently at $r<3$~kpc (galaxy profile) and $r>3$~kpc (ghostly halo profile). 

A galaxy disk is usually modeled as an exponential disk \citep{Kravtsov2013}, so our assumption is well justified. However, according to numerical simulations and literature for the Milky Way and other nearby massive galaxies, the functional form for the stellar halo profile is a power law density profile $\propto r^{-\alpha}$ with $\alpha \sim 4-5$ \citep{2006MNRAS.365..747A,Helmi2008}. In addition the stellar halo is not spherical but triaxial. However, over the limited range of radii where a stellar halo is observed in the dwarf galaxies (due to foreground contamination), the exponential fit and a power law fit are nearly identical. We choose to use an exponential profile to be consistent with our theoretical modeling assumption of homology, meaning we assumed the same density profile up to some scaling constants for the stellar halo and the disks/spheroids of the merging galaxies.

The total number of red-giant/AGB stars for the halo and the galaxy components were obtained integrating the two exponential fits from radii 0 to $4 r_{\rm h}^{\rm halo}$.
It is easy to show that for an exponential halo, $\Sigma(r) =\Sigma_0 e^{-r/r_{\rm exp}}$, the half-mass radius is $r_h=1.678 r_{\rm exp}$ and the number of red-giant (RG) stars is
\begin{align}
N_{\rm RG}(r) =2\pi \Sigma_0 r_{\rm exp}^2[1-(1+r/r_{\rm exp})e^{-r/r_{\rm exp}}].
\end{align}

Since the data is in terms of number of red-giant stars selected around the tip of the red-giant branch, we need to convert this observed number to a stellar mass at zero age main sequence (at the time of the galaxy formation) that in our case nearly coincides with the redshift $z=z_{\rm rei}$. We do this conversion using synthetic color-magnitude diagrams (CMDs), assuming stellar evolutionary tracks appropriate for dwarf galaxies that formed all their stars before reionization $Z=0.0003$ and age = 13~Gyr. Different references on the dwarf stellar halos we found in the literature use different filters, different magnitude and color cuts to separate the red-giants in the stellar halo of the dwarfs from the foreground M-stars in the Milky Way halo (the main contaminants dominating the foreground). Therefore, in Appendix~\ref{app:cmd}, we show our synthetic CMD for each of the six galaxies considered here, using the same cuts and filters as in the reference paper. In Table~\ref{tab:gal2} we show the final results of this procedure reporting, for each galaxy, the conversion factor, $M_{\rm conv}$, between number of stars and stellar mass at formation. Clearly, this is only relevant to calibrate the stellar mass of the halo and its surface brightness. The ratio of scale radii of the galaxy and the halo is independent of this calibration.
\begin{table*}
\begin{center}
    \begin{tabular}{llllllllll}
    \hline
    Galaxy & \multicolumn{3}{c}{Galactic Parameters} & \multicolumn{3}{c}{Ghostly Halo Parameters} & $r_{h}^{\rm halo}/r_h^{\rm Gal}$ & $M_{*}^{\rm halo}/M_{*}^{\rm Gal}$\\ 
    & $r_h^{\rm Gal} [kpc]$ & $M_{*, exp}^{Gal}$ [$M_{\odot}$] & $M_{*, mag}^{Gal}$ [$M_{\odot}$] & $r_h^{\rm halo} [kpc]$ & $M_{*, exp}^{\rm halo}$ [$M_{\odot}$] & $\Sigma^{\rm halo}_{h}$ [M$_\odot$ kpc$^{-2}$] \\ 
\hline
\\   
         Leo T & 0.09 & $3.2 \times 10^4$ & $2.80 \times 10^{4}$ & 0.30 & $2.6 \times 10^{4}$ & $4.6 \times 10^{4}$ & 3.3 & 0.82 \\
         
    Leo A & 0.31 & $3.0 \times 10^6$ & $2.80 \times 10^6$ & 0.87 & $7.8 \times 10^{5}$ & $1.6 \times 10^{5}$ & 2.8 & 0.26 \\
    
      WLM & 1.03 & $8.5 \times 10^6$ & $1.77 \times 10^7$ & 4.6 & $3.6 \times 10^{6}$ & $2.7 \times 10^{4}$ & 4.5 & 0.42 \\
      
       IC1613 & 1.16 & $1.19 \times 10^7$ & $3.69 \times 10^7$ & 4.9 & $4.4 \times 10^{6}$ & $2.8 \times 10^{4}$ & 4.3 & 0.37\\
       
       IC10 & 1.14 & $4.6 \times 10^7$ & $4.6 \times 10^7$ & 2.9 & $9.6 \times 10^{6}$ & $1.7 \times 10^{5}$ & 2.6 & 0.21 \\
       
      NGC6822 & 0.88 & $1.7 \times 10^7$ & $4.86 \times 10^7$ & 2.3 & $9.7 \times 10^{6}$ & $2.9 \times 10^{5}$ & 2.6 & 0.56\\
      
    \end{tabular}
    \caption{Galactic and ghostly halo parameters from fitting the observed stellar profiles and CMD (see Appendix~\ref{app:cmd}. The surface brightness of the ghostly halo within the half-mass radius is $\Sigma^{\rm halo}_{h}=M_{*, \rm exp}^{\rm halo}/[2\pi (r_h^{\rm halo})^2]$ and the half-mass radius is $r_{\rm h}=1.678r_{\rm exp}$. $M_{\rm *,exp}$ is the stellar mass obtained by integrating the exponential profile, while $M_{\rm *,mag}$ is the stellar mass obtained from the absolute magnitude assuming a constant mass-to-light ratio $\Upsilon =2$.}\label{tab:gal3}
    \end{center}
\end{table*}
In addition, we calculated the central galaxy stellar mass using the same procedure (integrating the galaxy exponential profile for the number of red-giant stars, and using the same conversion to stellar mass). However, the central galaxy is not composed of only old low-metallicity stars (for instance IC10 is a starburst galaxy, and NGC6822 has recent star formation). Therefore this estimate only picks the fraction of the mass in evolved stellar population and the stellar half-radius is also biased toward the old stellar population. Alternatively, for the central galaxy, we report the half-light radii from the literature \citep{McConnachie2012} and the stellar mass derived from the galaxy magnitude assuming a mass-to-light ratio $\Upsilon \equiv L_*/M_*=2$, with $\log{L_*} = -0.4(M_V - M_{V,sun})$, where $L_*$ and $M_*$ are in solar units and $M_V$ is the absolute magnitude taken from the literature (see Table~\ref{tab:gal1}). 
Table~\ref{tab:gal3} summarizes the galactic and stellar halo parameters for the six dwarfs considered here. These numbers will be the basis to constrain the star formation efficiency in the first galaxies using the procedure explained in the next section.
\begin{table*}
\begin{center}
    \begin{tabular}{lllllllllllll}
    \hline
    Galaxy & $M_{dyn}(\leq r_h)^{(a)}$ & $M_{1/2}^{(b)}$ & BD15(NFW)$^{(c)}$ & BD15(DC14)$^{(c)}$ & $M_{200}^{(d)}$ & $M_{\rm dm}(M_*)$ & $M_{\rm dm}(r_{\rm h})$& Fiducial\\
    & [$\rm M_{\odot}$] & [$\rm M_{\odot}$] & [$\rm M_{\odot}$] & [$\rm M_{\odot}$] & [$\rm M_{\odot}$] & [$\rm M_{\odot}$] & [$\rm M_{\odot}$] & [$\rm M_{\odot}$] \\
\hline
\\
   Leo T & $3.9 \times 10^6$ &$6.0 \times 10^6$ & $9.05 \times 10^9$ & $9.05 \times 10^9$ & $3.5-7.5 \times 10^8$ & $1.76\times 10^8$ & $2.27\times 10^7$ (0.17) & $2.4 \times 10^8$\\
   
   Leo A & $2.5 \times 10^7$ & $1.5 \times 10^7$ & $3.00 \times 10^8$ & $8.01 \times 10^9$ & - & $2.96\times 10^9$ & $2.96\times 10^8$ (0.4) & $1.6 \times 10^9$\\
    
      WLM & $3.80 \times 10^8$ & $6.30 \times 10^8$ & $6.02 \times 10^9$ & $1.58 \times 10^{10}$ & $8.3 \times 10^9$ & $5.83\times 10^9$ & $1.24\times 10^{10}$ (1.39)& $8.8 \times 10^9$\\
      
       IC1613 & $1.7 \times 10^{9}$ & - & $1.10 \times 10^8$ & $2.44 \times 10^9$ & $1.7 \times 10^{9}$ & $7.89\times 10^9$ & $1.1\times 10^{10}$ (1.33)& $6.9 \times 10^{9}$\\
        IC10 & - & - & - & - & - & $1.98\times 10^{10}$ & $7.41\times 10^9$ (1.17)& $1.4 \times 10^{10}$\\
        
      NGC6822 & - & $2.40 \times 10^8$ & - & - & $2 \times 10^{10}$ & $1.04\times 10^{10}$ & $1.38\times 10^{10}$ (1.44) & $1.5 \times 10^{10}$\\    
    \end{tabular}
    \caption{Various estimates of the of dark matter halo mass of the dwarf galaxies in Table~\ref{tab:gal1}. Second to fifth columns represent estimated masses from literatures. (a) \citet{McConnachie2012}; (b) \citet{Mdm2014}; (c) \citet{Mdm2015}; (d) \citet{Mdm2017}; Blank indicates lack of data. See text for more explanation.}\label{tab:dm}
    \end{center}    
\end{table*}

Estimates of the total dark matter halo mass of the observed dwarfs are shown in Table~\ref{tab:dm}. This table collects, for each galaxy, the values from three main methods we used to estimate the masses. First, we report estimates found in the literature, when available. The discrepancy between different authors can be up to one order of magnitude. The dynamical mass of the halos $M_{\rm dyn}(\leq r_h)$ and the mass within the half light radius $M_{1/2}$, when available, are reported for completeness and provide a robust lower limit to the total mass, but are not used to estimate fiducial dark matter halo masses. The second method to estimate the total halo mass is based on the knowledge of the stellar mass of the galaxy at $z=0$. We use the relationship $M_*=6 \times 10^5~{\rm M}_\odot (M_{dm}/10^9~{\rm M}_\odot)^{1.5}$ obtained from extrapolating to lower masses the results from \cite{Behroozi2013} at $z=0$ to estimate $M_{\rm dm}(M_*)$ shown in the table.
The third method uses the relationship between the half-light radius and the halo mass in Equation~(\ref{eq:rh}) to obtain $M_{\rm dm}(r_{\rm h})$, using for $r_{\rm h}$ the results from the exponential profile fits in Table~\ref{tab:gal3}. The last two methods may have an uncertainty of up to a factor of 3, but they seem to provide estimates consistent with each other. Finally, the last column in the table shows the fiducial mass of the dark matter halo obtained from the average of three different methods: the value from the literature in BD15, $M_{\rm dm}(M_*)$ and $M_{\rm dm}(r_{\rm h})$.

\section{Star Formation Efficiency in Fossil Galaxies}\label{sec:constr}

Given the observational data in \S~\ref{sec:obs}, we use the theoretical modeling in \S~\ref{sec:res} to infer the values of $\beta$ and $\epsilon_0$, and therefore the star formation efficiency $f_*(M_{\rm dm})$ at $z \sim 6$. This procedure, explained in detail below, is done independently for each one of the six dwarf galaxies in Table~\ref{tab:gal1}. We also explore the sensitivity of the results to the uncertainties in estimating $M_{\rm dm,gal}$, $M_{\rm res}$ and $z_{\rm cut}$.

\begin{figure*}
\includegraphics[width=0.49\textwidth]{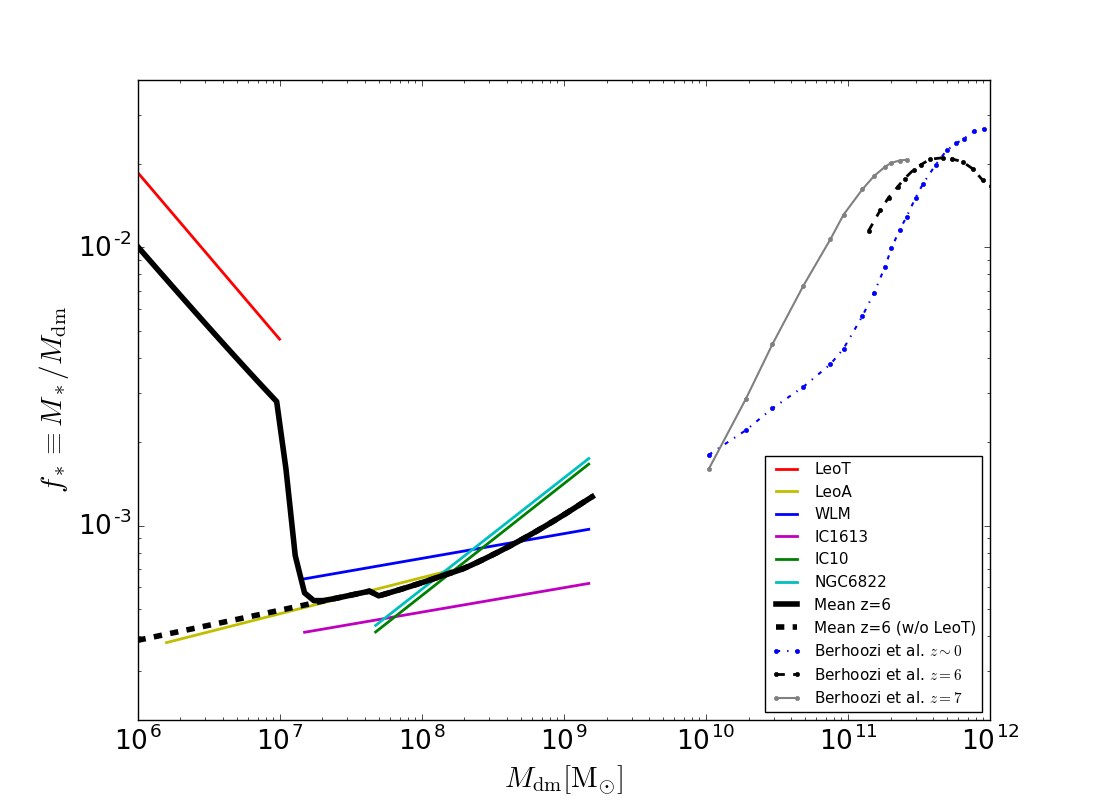}
\includegraphics[width=0.49\textwidth]{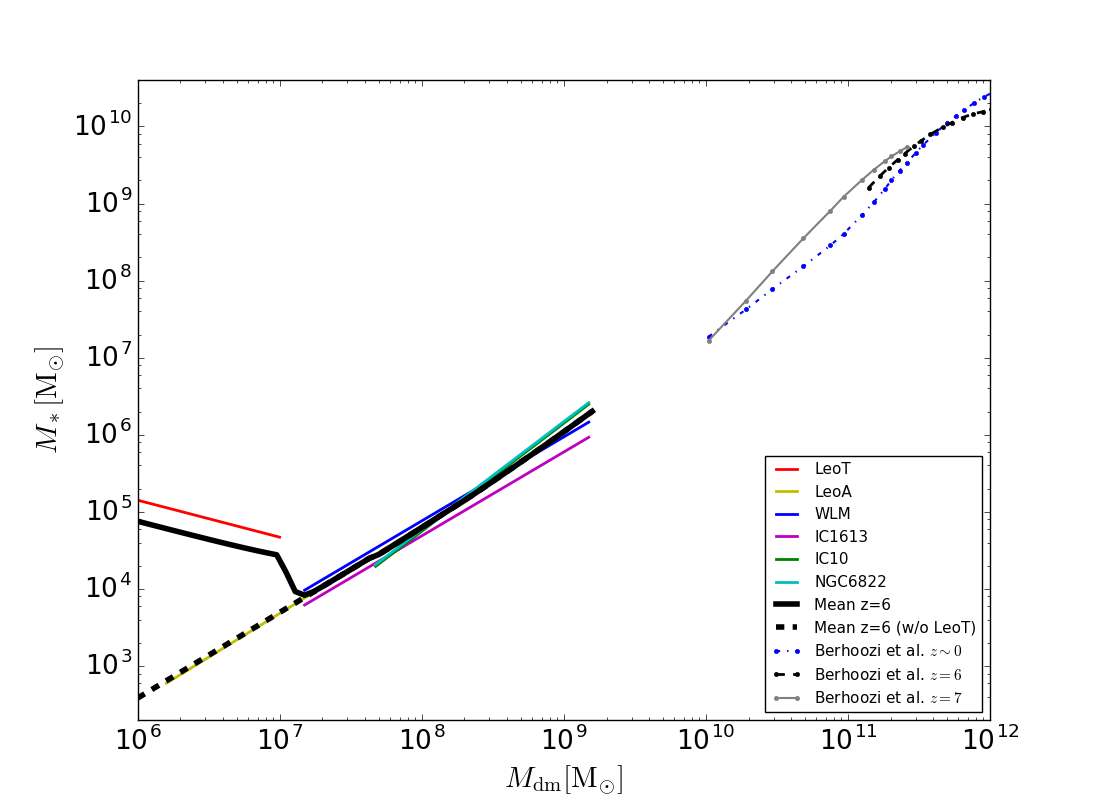}
\caption{{\it (Left).} Star formation efficiency as a function of dark matter halo mass at $z=6$ derived from data on ghostly stellar halos in dwarf galaxies and using the fiducial parameters for the stellar halo model (this paper). The thin lines refer to the results from individual dwarf galaxies as in the legend. The thick black line is the mean value considering all the dwarf galaxies (solid line) and excluding Leo~T (dashed line). For comparison we show the results from \citep{Behroozi2013}, valid for masses $>10^{10}$~M$_\odot$ and at redshifts $z=6,7$ and $z=0$ as shown in the legend. {\it (Right).} Same as in the left panel but showing $M_*$ as a function of the dark matter halo mass.}\label{fig:res1}
\end{figure*}

We first constrain $\beta$ from the ratio $r_{\rm h}^{\rm halo}/r_{\rm h}^{\rm gal}$ since this ratio depends only on $\beta$ and the dark matter halo mass. We compare the model results in Figure~\ref{fig:radius}(bottom) to the observed values in Table~\ref{tab:gal3} for each dwarf galaxy separately. 
Next, we constrain the normalization, $\epsilon_0$, by comparing the observed $M_*^{\rm halo}$ to the model results in Figure~\ref{fig:mass}(top). Inspecting Figure~\ref{fig:mass}(top) we notice that when using a pivot mass, $M_{dm,0}=10^7$~M$_\odot$, to normalize the star formation efficiency, the stellar halo mass has a weak dependence on the value of $\beta$. Therefore the main uncertainty on $\epsilon_0$ is the estimated dark matter halo mass rather than the uncertainty on $\beta$. Given $\beta$ and $\epsilon_0$, we plot $f_*(M_{\rm dm})$ and $M_*(M_{\rm dm}) \equiv f_* M_{\rm dm}$ as a function of $M_{\rm dm}$ for each one of the six dwarfs in our sample as shown by the solid thin lines in Figure~\ref{fig:res1}. The lines are plotted only for a  range of dark matter halo masses bounded by an estimate of the maximum mass of merging satellites and $1/100$ of this mass. The solid thick lines show the mean values of $f_*$ and $M_*$ considering all the dwarfs, while the dashed thick line is the same but excluding Leo~T from the estimate of the mean.
Here we have used our fiducial values for $M_{\rm dm,gal}$ (see Table~\ref{tab:dm}), $M_{\rm res}=10^6$~$M_\odot$ and $z_{\rm cut}=6$.
For comparison we also show the results obtained by \cite{Behroozi2013} for much more massive halos at $z=0$, $z=6$ and $z=7$ as shown by the lines with data points (see the legend).

The results from individual dwarfs are remarkably consistent with each other and with a naive extrapolation of \cite{Behroozi2013} result. Given the significant uncertainties on $M_{\rm dm,gal}$, given the intrinsic scatter of the theoretical results due to different merging histories, and considering the observational errors on $r_{\rm h}^{\rm halo}/r_{\rm h}^{\rm gal}$ and $M_*^{\rm Halo}$, this result is very encouraging and gives us confidence that this method can be significantly improved with better observational data and modeling.

The only dwarf galaxy that produced a result that deviates significantly from the others is Leo~T, which is also the least massive dwarf galaxy in the our sample. The results for Leo~T are consistent with a significant increase of $f_*$ for the mass range $M_{\rm dm} \sim 10^6-10^7$~M$_\odot$. The results for Leo~T are not consistent with the other dwarfs because are valid for halos with $M_{\rm dm,gal}<10^7$~M$_\odot$, which have a negligible impact on the properties of ghostly halos in the other dwarfs which are significantly more massive. If the results for Leo~T are correct, the star formation efficiency appears to increase sharply from $f_*=0.1\%$ at $M_{\rm dm}\sim 10^7$~M$_\odot$ to $f_*=5\%$ at $M_{\rm dm}\sim 10^6$~M$_\odot$.
This result is very intriguing but tentative. The results of some numerical  simulations of the first galaxies are consistent with a sharp increase of $f_*$ in the first small mass halos. However, given that we only have one object showing this trend and given the large uncertainties of the data and modeling, we remain cautious to avoid over-interpreting this result.

\begin{figure*}
\includegraphics[width=0.48\textwidth]{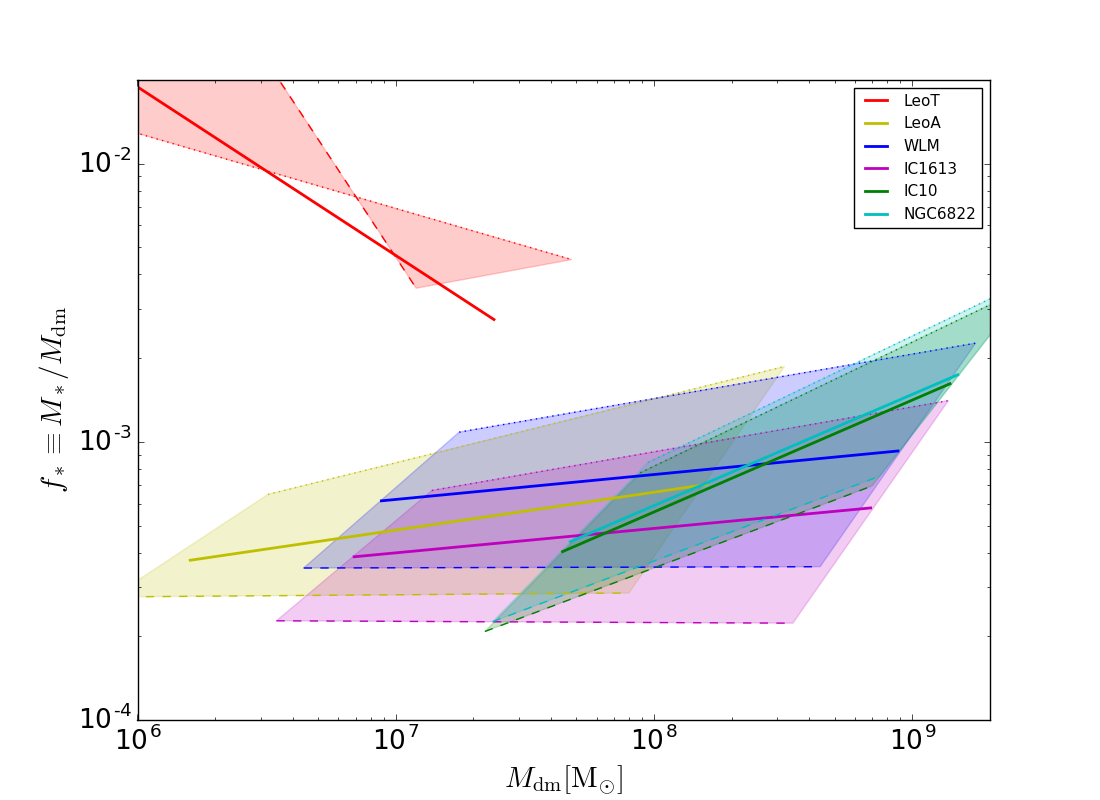}
\includegraphics[width=0.48\textwidth]{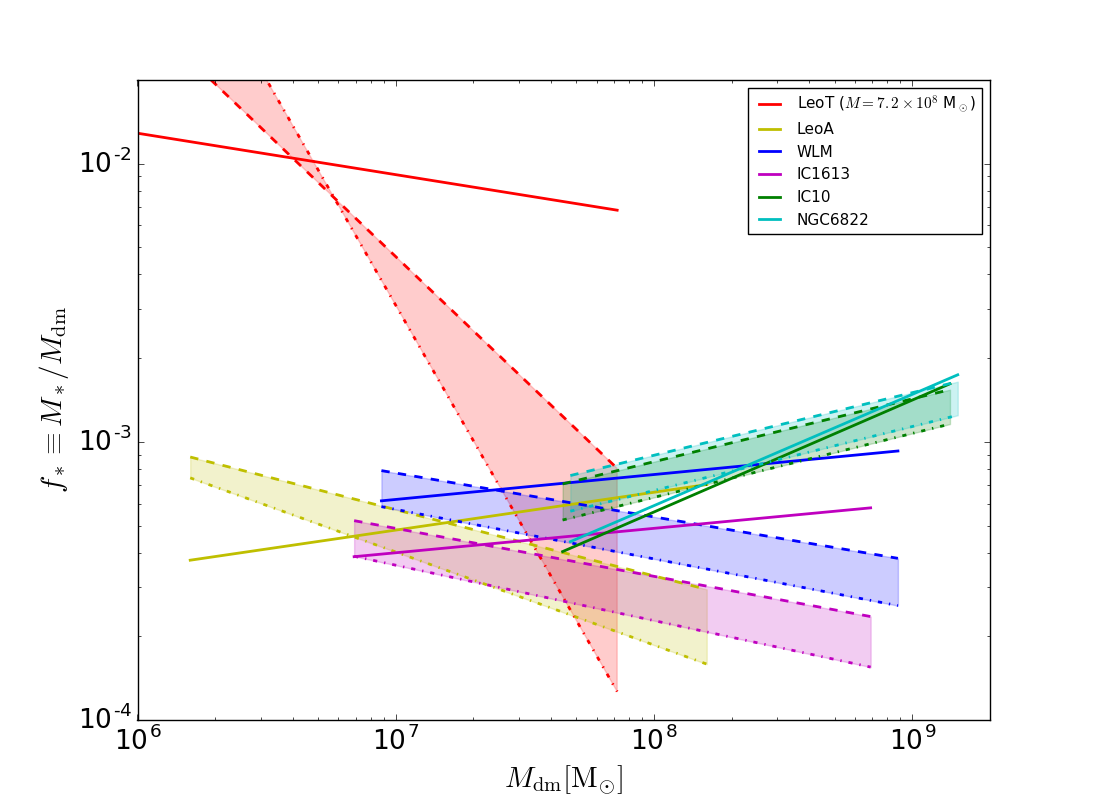}
\caption{{\it (Left.)} Same as Fig.~\ref{fig:res1}(left) but showing the sensitivity of the the results to difference choices for the dark matter halo mass of the dwarf galaxy. The area shown by the shaded region for each dwarf shows the fiducial model (thick solid line) and $2\times$ (thin solid line) and $1/2 \times$ (thin dashed line) the value of the dark matter halo mass. The overall normalization of $f_*$ ({\it i.e.}, the value of $\epsilon_0$) depends nearly linearly on the assumed dark matter halo mass of the dwarfs, while the slope of $f_*$ ({\it i.e.}, $\beta$) remains nearly unchanged. {\it (Right).} Same as the left panel but showing the sensitivity of the results to $z_{\rm cut}$ and $M_{\rm res}$. The fiducial values are shown as solid think lines, while the thin dashed line assumes $M_{\rm res}=5 \times 10^6$~M$_\odot$ with $z_{\rm cut}=6$ and the thin dotted line assumes $M_{\rm res}=5 \times 10^6$~M$_\odot$ with $z_{\rm res}=5$. The values of $\beta$ (slope of the lines for $f_*$) are rather sensitive to $M_{\rm res}$ while the dependence on $z_{\rm cut}$ is weaker.}\label{fig:res2}
\end{figure*}

Given the difficulty in obtaining a precise estimate of $M_{dm,gal}$ for each dwarf galaxy, in Figure~\ref{fig:res2}(left) we show the effect on $f_*$ of doubling or reducing by half its value with respect to the fiducial value in Table~\ref{tab:dm}. The fiducial results for $f_*$ are show as solid lines, while the shaded region is bounded at the top by twice the fiducial mass and at the bottom by half the fiducial mass. 
 The right panel in Figure~\ref{fig:res2} is similar to the left panel but for $M_{\rm res}=5 \times 10^6$~M$_\odot$ with $z_{\rm cut}=6$ (dashed lines) and with $z_{\rm cut}=5$ (dot-dashed lines). The solid lines refer to the fiducial case as in the left panel. From the figure it is evident that $z_{\rm cut}$ has a small effect on the results while $M_{\rm res}$ has a significant effect on $\beta$. Instead $M_{\rm dm,gal}$ has a strong effect on $\epsilon_0$. Note that for Leo~T it is not possible to find a model reproducing the observations if the cutoff mass of luminous galaxies is $M_{\rm res}=5 \times 10^6$~M$_\odot$, for the fiducial value of Leo~T mass. A solution is found in this case only assuming that the dark matter mass of Leo~T is at least 3 times larger than the fiducial value. 
The modeling results for $z_{\rm cut}=6$ show that for the values of $\beta$ obtained in this study, which are in the range $\beta \sim 0 - 0.5$, the surface mass density is relatively constant as a function of the halo mass and is roughly between $\Sigma_*^{\rm halo}\sim 3\times 10^4 (\epsilon_0/0.001)$ and $10^5 (\epsilon_0/0.001)$. Table~\ref{tab:gal3} shows that the observed values of $\Sigma_*^{\rm halo}$ are in the same range and with the correct dependence on the halo mass (with the exception of Leo~A).

As shown in Figure~\ref{fig:res1}, if we exclude Leo~T from the sample, the stellar halo of the remaining dwarfs is consistent with a star formation efficiency in the first galaxies at $z\sim 6$ nearly independent of the halo mass, meaning that roughly $M_* \propto M_{\rm dm}$. The star formation efficiency in halos of $10^6-10^8$~M$_\odot$ is rather large considering the small masses of these halos: $f_* \sim 0.05\%-0.1 \%$. Therefore these halos host galaxies with stellar masses, $M_* \sim 5 \times 10^2~{\rm M}_\odot (M_{\rm dm}/10^6~{\rm M}_\odot)$, at formation. These stellar masses are typical of UFDs satellites observed in the Local Group. The result is consistent with the identification of UFDs as fossil of the first galaxies, but implies that the faintest UFDs that have been observed so far, with $M_* \sim 10^3$~M$_\odot$, have formed in dark matter halos with masses $\sim 10^7$~M$_\odot$ at $z>6$, and reside in $\sim 10^8$~M$_\odot$ halos today. This also implies that UFDs with $M_* \sim 10^3-10^4$~M$_\odot$ should be very numerous because they are hosted in the smallest mass halos that can form Pop~II stars\footnote{Minihalos with masses $M_{\rm dm}\le 10^{6}$~M$_\odot$ are believed to host Pop~III stars with a top-heavy IMF and therefore do not leave visible fossil remnants}, which are the most numerous in CDM cosmologies. 

To summarize, our results constrain the star formation efficiency in the smallest luminous galactic building blocks with halo masses $10^6-10^9$~M$_\odot$. When comparing our results to modeling by \cite{Behroozi2013}, valid at the same redshift but for significantly more massive galaxies ($M_{dm}>10^{10}$~M$_\odot$), a consistent picture seems to emerge in which feedback effect are less important in reducing the star formation efficiency below a redshift-dependent critical mass for dark matter halos.

\section{Summary and Discussion}\label{sec:sum}

We carry out numerical simulations to characterize the size, stellar mass, and stellar mass surface density of extended stellar halos in dwarf galaxies as a function of dark matter halo mass. We expect that for galaxies smaller than a critical value, these ghostly halos will not exist because the smaller galactic subunits that build it up do not form any stars.  However, if a stellar halo is found in sufficiently small mass dwarfs, the whole stellar halo is composed of tidal debris of fossil galaxies, {\it i.e.,} galaxies that formed most of their stars before the epoch of reionization. We refer to these stellar halos in dwarf galaxies as ghostly halos. Therefore the detection and characterization of ghostly stellar halos around isolated dwarf galaxies is a sensitive test of the efficiency of star formation in fossil galaxies. The properties of ghostly halos are tightly connected to the properties of the population of ultra-faint dwarf galaxies found around the Milky Way and M31. This elusive galaxy population still needs to be fully uncovered and understood and is one of the most powerful probes of the physics of galaxy formation in the early universe and the epoch of reionization. 
 
We have characterized stellar halos in dwarf galaxies as a function of their present-day dark matter halo mass and as a function of a parametrization of the star formation efficiency at $z=z_{\rm cut}\sim 6$, approximated by a power-law $\propto M_{\rm dm}^\beta$. From a literature search we found that 6 of the 12 dwarf galaxies in the Local Group residing outside the virial radii of the Milky Way and M31 show some evidence of an extended stellar halo. We analyzed the properties of these stellar halos in a uniform way and compared their properties to the expectations from the theoretical modeling. From the observational data and its comparison to the models emerged a rather consistent picture that allowed us to constrain the star formation efficiency at $z\sim 6$. We find that at redshift $z\sim 6$, dark matter halos in the mass range $10^7-10^9$~M$_\odot$ have a mean star formation efficiency in the range $f_* \equiv M_*/M_{dm} \sim 0.05\%-0.1\%$, only mildly increasing as a function of the dark matter halo mass ($\beta \sim 0.1$). The result extends by several orders of magnitude to lower halo masses of the results from the literature valid for $M_{\rm dm}>10^{10}$~M$_\odot$ \citep{Behroozi2013}. 

The main implications and the interpretation of the results found in this study are summarize below:
\begin{enumerate}
\item Star formation efficiency nearly constant as a function of $M_{\rm dm}$ has implications on the role of feedback in regulating star formation in the smallest mass halos. Typically small halo masses are less bound gravitationally and more susceptible to the effect of SNe and radiation feedbacks; thus the result is somewhat counter-intuitive. However, it is not completely unexpected. Observations of old globular clusters (GCs) \citep{McLaughlin:99,GeorgievP2010}, theoretical modeling the Milky Way system of GCs \citep{KatzR2014}, and constraint on the formation epoch of old GCs using deep field luminosity functions \citep{Ricotti2002, KatzR2013}, suggest that the star formation efficiency of old GCs is independent of the halo mass in dwarf galaxies. In \cite{RicottiPG2016} we suggested a common origin for GCs and UFDs: namely, star formation in the first galaxies always happens in compact star clusters, some of which become unbound and evaporate in the first small mass halos and transform into UFDs. Regarding the role of feedback, the result can be interpreted as follows: below a critical stellar mass in dwarf galaxies, the "quantization" in compact star clusters becomes dominant. The minimum cluster mass is regulated by stellar cluster feedback which operates on shorter time-scales than galaxy scale feedback. UFDs with total stellar masses of $10^3-10^4$~M$_\odot$ are at most composed of a few stellar clusters forming in a nearly instantaneous mode of star formation. Galaxy scale feedback starts operating as a result of the formation of the first few clusters, but the clusters themselves have a characteristic mass scale and therefore provide a minimum stellar mass floor.
\item From the mean star formation efficiency we can make a simple estimate of the faint end of the luminosity function before reionization. If the faintest galactic building blocks have stellar masses of $10^3$~M$_\odot$, their peak luminosity, assuming an instantaneous ($<5$~Myr) burst of star formation, has an absolute magnitude $M_V \sim -8$. So, the faint end of the galaxy luminosity function at $z>6$ may become flatter due to the short duty cycle of the burst, but will have a faint cutoff at about this magnitude. This is largely consistent with the result of numerical simulations \citep{RicottiGS2002b,RicottiGS2008,Wise2012,RicottiPG2016}.
\item  According to our analysis and results, galaxies with luminosities typical of UFDs are hosted in small mass halos of $10^6-10^8$~M$_\odot$ at $z=6$, which end up today in very numerous low-mass galactic satellites. We thus expect the number of yet undiscovered UFDs in the Milky Way and M31 to be large.
\item The tentative existence of an extended stellar halo in the faintest galaxy in our sample, Leo~T, suggests a significant increase of $f_*$ for the mass range $M_{\rm dm} \sim 10^6-10^7$~M$_\odot$.
If the results for Leo~T are correct, the star formation efficiency appears to increase sharply from $f_*=0.1\%$ at $M_{\rm dm}\sim 10^7$~M$_\odot$ to $f_*=5\%$ at $M_{\rm dm}\sim 10^6$~M$_\odot$.
This result is intriguing but tentative. More and better data and further theoretical modeling will be necessary to put this result on firmer ground. 
\end{enumerate}
This work is meant as a first step and a proof of concept of a new method to constrain star formation before reionization. We have made several simplifying assumptions in the theoretical modeling that need to be checked with numerical simulations and that will be the subject of future work. One important assumption we need to relax is the one-to-one relationship between halo mass and stellar mass. Simulations show that in the mass range $10^6-10^8$~M$_\odot$, the scatter around a mean star formation efficiency is very large \citep[{\it e.g.},][]{RicottiGS2002b}. The effects of this stochasticity in the mass-to-light ratio in the first galaxies needs to be included in future models to check whether it has an effect on the scale radius of ghostly halos. 

Another front in which a lot of progress can be made is on improving and expanding the observational data. Deeper observations of the six isolated dwarf galaxies identified in this paper can help improve the characterization of their stellar halos and check for possible contamination with higher metallicity stars produced by processes other than mergers with fossil dwarfs, {\it e.g.,} breathing or evaporation of disk stars or complex dynamical interactions with massive companions \citep{2009ApJ...702.1058Z,2014MNRAS.440.1971W,2018Natur.555..334B}.

Perhaps more important and exciting implication is the possibility to discover new extended stellar halos around isolated dwarfs. Looking at the census of known dwarfs in the Local Group, we noticed some interesting statistics. 

Leo~T and NGC~6822, which both belong to our list because show tentative evidence of an extended stellar halo, are the only two dwarfs belonging to the Milky Way sub-group and residing outside its virial radius \citep{McConnachie2012}. Phoenix dIrr ($M_*=7.7\times 10^5$~M$_\odot$, $r_h=454$~pc) also belongs to this group but its radial velocity appears marginally larger than the Milky Way escape velocity, thus may not be bound to the Milky Way. Nevertheless it is a good candidate for future studies.

Of the four dwarfs belonging to M31 sub-group but outside its virial radius, two, IC10 and IC1613, are in our list. The other two are LGS3 ($M_*=9.6\times 10^5$~M$_\odot$, $r_h=470$~pc) and AndVI ($M_*=2.8\times 10^6$~M$_\odot$, $r_h=524$~pc). PegDIG ($M_*=6.6\times 10^6$~M$_\odot$, $r_h=562$~pc) also belongs to this group but its radial velocity appears marginally larger than M31 escape velocity \citep{McConnachie2012}.

Finally, of the six isolated dwarfs belonging to the Local Group, two, Leo~A and WLM, belong to our list. The other four are Aquarius  ($M_*=1.6\times 10^6$~M$_\odot$, $r_h=458$pc), Tucana ($M_*=5.6\times 10^5$~M$_\odot$, $r_h=284$~pc), SagDIG ($M_*=3.5\times 10^6$~M$_\odot$, $r_h=282$~pc), UGC4879 ($M_*=8.3\times 10^6$~M$_\odot$, $r_h=162$~pc). These dwarfs are either dIrr or transitional galaxies dIrr/dSph, and are rather faint with sizes and stellar masses similar to Leo~T and Leo~A, which are the faintest dwarfs in our list. We have checked the literature for evidence of an extended halo in these galaxies but we did not find any paper showing its existence. The only exception is SagDIG, for which we have found a published radial profile \citep{SagDIG2016}, but it did not show evidence for two stellar components (galaxy and halo) and the galaxy showed a possible signature of tidal stripping. 

It may be challenging to detect extended stellar halos in these dwarfs, if they exist, given their faintness and large distance from the Milky Way ($\sim 1-1.5$~Mpc). However, we suggest this list of dwarfs as a promising target for future observations.

\subsection*{ACKNOWLEDGMENTS}
MR thank the National Science Foundation for support under the
Theoretical and Computational Astrophysics Network (TCAN) grant
AST1333514 and CDI grant CMMI1125285. Thanks to the anonymous
referee.




\bibliographystyle{mnras}
\bibliography{References} 

\begin{thebibliography}{}
\makeatletter
\relax
\def\mn@urlcharsother{\let\do\@makeother \do\$\do\&\do\#\do\^\do\_\do\%\do\~}
\def\mn@doi{\begingroup\mn@urlcharsother \@ifnextchar [ {\mn@doi@}
  {\mn@doi@[]}}
\def\mn@doi@[#1]#2{\def\@tempa{#1}\ifx\@tempa\@empty \href
  {http://dx.doi.org/#2} {doi:#2}\else \href {http://dx.doi.org/#2} {#1}\fi
  \endgroup}
\def\mn@eprint#1#2{\mn@eprint@#1:#2::\@nil}
\def\mn@eprint@arXiv#1{\href {http://arxiv.org/abs/#1} {{\tt arXiv:#1}}}
\def\mn@eprint@dblp#1{\href {http://dblp.uni-trier.de/rec/bibtex/#1.xml}
  {dblp:#1}}
\def\mn@eprint@#1:#2:#3:#4\@nil{\def\@tempa {#1}\def\@tempb {#2}\def\@tempc
  {#3}\ifx \@tempc \@empty \let \@tempc \@tempb \let \@tempb \@tempa \fi \ifx
  \@tempb \@empty \def\@tempb {arXiv}\fi \@ifundefined
  {mn@eprint@\@tempb}{\@tempb:\@tempc}{\expandafter \expandafter \csname
  mn@eprint@\@tempb\endcsname \expandafter{\@tempc}}}

\bibitem[\protect\citeauthoryear{{Abadi}, {Navarro}  \& {Steinmetz}}{{Abadi}
  et~al.}{2006}]{2006MNRAS.365..747A}
{Abadi} M.~G.,  {Navarro} J.~F.,   {Steinmetz} M.,  2006, \mn@doi [\mnras]
  {10.1111/j.1365-2966.2005.09789.x}, \href
  {http://adsabs.harvard.edu/abs/2006MNRAS.365..747A} {365, 747}

\bibitem[\protect\citeauthoryear{{Babul} \& {Rees}}{{Babul} \&
  {Rees}}{1992}]{Babul92}
{Babul} A.,  {Rees} M.~J.,  1992, \mn@doi [\mnras]
  {10.3847/0004-637X/831/2/204}, \href {xx} {255, 346}

\bibitem[\protect\citeauthoryear{{Battinelli} \& {Demers}}{{Battinelli} \&
  {Demers}}{2009}]{Battinelli2009}
{Battinelli} P.,  {Demers} S.,  2009, \mn@doi [\aap]
  {10.1051/0004-6361:200810619}, \href
  {http://adsabs.harvard.edu/abs/2009A&A...493.1075B} {493, 1075}

\bibitem[\protect\citeauthoryear{{Battinelli}, {Demers}  \&
  {Kunkel}}{{Battinelli} et~al.}{2006}]{2006NGC6822}
{Battinelli} P.,  {Demers} S.,   {Kunkel} W.~E.,  2006, \mn@doi [\aap]
  {10.1051/0004-6361:20054718}, \href
  {http://adsabs.harvard.edu/abs/2006AAp...451...99B} {451, 99}

\bibitem[\protect\citeauthoryear{{Behroozi}, {Wechsler}  \&
  {Conroy}}{{Behroozi} et~al.}{2013}]{Behroozi2013}
{Behroozi} P.~S.,  {Wechsler} R.~H.,   {Conroy} C.,  2013, \mn@doi [\apj]
  {10.1088/0004-637X/770/1/57}, \href
  {http://adsabs.harvard.edu/abs/2013ApJ...770...57B} {770, 57}

\bibitem[\protect\citeauthoryear{{Bergemann} et~al.,}{{Bergemann}
  et~al.}{2018}]{2018Natur.555..334B}
{Bergemann} M.,  et~al., 2018, \mn@doi [\nat] {10.1038/nature25490}, \href
  {http://adsabs.harvard.edu/abs/2018Natur.555..334B} {555, 334}

\bibitem[\protect\citeauthoryear{{Bovill} \& {Ricotti}}{{Bovill} \&
  {Ricotti}}{2009}]{BovillR2009}
{Bovill} M.~S.,  {Ricotti} M.,  2009, \mn@doi [\apj]
  {10.1088/0004-637X/693/2/1859}, \href
  {http://adsabs.harvard.edu/abs/2009ApJ...693.1859B} {693, 1859}

\bibitem[\protect\citeauthoryear{{Bovill} \& {Ricotti}}{{Bovill} \&
  {Ricotti}}{2011a}]{BovillR2011a}
{Bovill} M.~S.,  {Ricotti} M.,  2011a, \mn@doi [\apj]
  {10.1088/0004-637X/741/1/17}, \href
  {http://adsabs.harvard.edu/abs/2011ApJ...741...17B} {741, 17}

\bibitem[\protect\citeauthoryear{{Bovill} \& {Ricotti}}{{Bovill} \&
  {Ricotti}}{2011b}]{BovillR2011b}
{Bovill} M.~S.,  {Ricotti} M.,  2011b, \mn@doi [\apj]
  {10.1088/0004-637X/741/1/18}, \href
  {http://adsabs.harvard.edu/abs/2011ApJ...741...18B} {741, 18}

\bibitem[\protect\citeauthoryear{{Boylan-Kolchin}, {Ma}  \&
  {Quataert}}{{Boylan-Kolchin} et~al.}{2005}]{BK2005}
{Boylan-Kolchin} M.,  {Ma} C.-P.,   {Quataert} E.,  2005, \mn@doi [\rm{MNRAS}]
  {10.1111/j.1365-2966.2005.09278.x}, \href
  {http://adsabs.harvard.edu/abs/2005MNRAS.362..184B} {362, 184}

\bibitem[\protect\citeauthoryear{{Boylan-Kolchin}, {Ma}  \&
  {Quataert}}{{Boylan-Kolchin} et~al.}{2006}]{BK2006}
{Boylan-Kolchin} M.,  {Ma} C.-P.,   {Quataert} E.,  2006, \mn@doi [\rm{MNRAS}]
  {10.1111/j.1365-2966.2006.10379.x}, \href
  {http://adsabs.harvard.edu/abs/2006MNRAS.369.1081B} {369, 1081}

\bibitem[\protect\citeauthoryear{{Brocato}, {Castellani}, {Raimondo}  \&
  {Romaniello}}{{Brocato} et~al.}{1999}]{Brocato1999}
{Brocato} E.,  {Castellani} V.,  {Raimondo} G.,   {Romaniello} M.,  1999,
  \mn@doi [\aaps] {10.1051/aas:1999198}, \href
  {http://adsabs.harvard.edu/abs/1999A&AS..136...65B} {136, 65}

\bibitem[\protect\citeauthoryear{{Brocato}, {Castellani}, {Poli}  \&
  {Raimondo}}{{Brocato} et~al.}{2000}]{Brocato2000}
{Brocato} E.,  {Castellani} V.,  {Poli} F.~M.,   {Raimondo} G.,  2000, \mn@doi
  [\aaps] {10.1051/aas:2000357}, \href
  {http://adsabs.harvard.edu/abs/2000A&AS..146...91B} {146, 91}

\bibitem[\protect\citeauthoryear{{Brook} \& {Di Cintio}}{{Brook} \& {Di
  Cintio}}{2015}]{Mdm2015}
{Brook} C.~B.,  {Di Cintio} A.,  2015, \mn@doi [\mnras] {10.1093/mnras/stv864},
  \href {http://adsabs.harvard.edu/abs/2015MNRAS.450.3920B} {450, 3920}

\bibitem[\protect\citeauthoryear{{Bullock}, {Kravtsov}  \&
  {Weinberg}}{{Bullock} et~al.}{2000}]{Bullock00}
{Bullock} J.~S.,  {Kravtsov} A.~V.,   {Weinberg} D.~H.,  2000, \mn@doi [\apj]
  {10.1086/309279}, \href {http://adsabs.harvard.edu/abs/2000ApJ...539..517B}
  {539, 517}

\bibitem[\protect\citeauthoryear{{Cantiello}, {Raimondo}, {Brocato}  \&
  {Capaccioli}}{{Cantiello} et~al.}{2003}]{Cantiello2003}
{Cantiello} M.,  {Raimondo} G.,  {Brocato} E.,   {Capaccioli} M.,  2003,
  \mn@doi [\aj] {10.1086/375322}, \href
  {http://adsabs.harvard.edu/abs/2003AJ....125.2783C} {125, 2783}

\bibitem[\protect\citeauthoryear{{Efstathiou}}{{Efstathiou}}{1992}]{Efstathiou92b}
{Efstathiou} G.,  1992, \mn@doi [\mnras] {10.3847/0004-637X/831/2/204}, \href
  {xx} {256, 43P}

\bibitem[\protect\citeauthoryear{{Eggen}, {Lynden-Bell}  \& {Sandage}}{{Eggen}
  et~al.}{1962}]{Eggen1962}
{Eggen} O.~J.,  {Lynden-Bell} D.,   {Sandage} A.~R.,  1962, \mn@doi [\apj]
  {10.1086/147433}, \href {http://adsabs.harvard.edu/abs/1962ApJ...136..748E}
  {136, 748}

\bibitem[\protect\citeauthoryear{{Georgiev}, {Puzia}, {Goudfrooij}  \&
  {Hilker}}{{Georgiev} et~al.}{2010}]{GeorgievP2010}
{Georgiev} I.~Y.,  {Puzia} T.~H.,  {Goudfrooij} P.,   {Hilker} M.,  2010,
  \mn@doi [\mnras] {10.1111/j.1365-2966.2010.16802.x}, \href
  {http://adsabs.harvard.edu/abs/2010MNRAS.406.1967G} {406, 1967}

\bibitem[\protect\citeauthoryear{{Gerbrandt}, {McConnachie}  \&
  {Irwin}}{{Gerbrandt} et~al.}{2015}]{2015IC10}
{Gerbrandt} S.~A.~N.,  {McConnachie} A.~W.,   {Irwin} M.,  2015, \mn@doi
  [\mnras] {10.1093/mnras/stv2029}, \href
  {http://adsabs.harvard.edu/abs/2015MNRAS.454.1000G} {454, 1000}

\bibitem[\protect\citeauthoryear{{Gnedin}}{{Gnedin}}{2000}]{Gnedin00b}
{Gnedin} N.~Y.,  2000, \mn@doi [\apj] {10.3847/0004-637X/831/2/204}, \href {xx}
  {542, 535}

\bibitem[\protect\citeauthoryear{{Helmi}}{{Helmi}}{2008}]{Helmi2008}
{Helmi} A.,  2008, \mn@doi [\aapr] {10.1007/s00159-008-0009-6}, \href
  {http://adsabs.harvard.edu/abs/2008A&ARv..15..145H} {15, 145}

\bibitem[\protect\citeauthoryear{{Higgs} et~al.,}{{Higgs}
  et~al.}{2016}]{SagDIG2016}
{Higgs} C.~R.,  et~al., 2016, \mn@doi [\mnras] {10.1093/mnras/stw257}, \href
  {http://adsabs.harvard.edu/abs/2016MNRAS.458.1678H} {458, 1678}

\bibitem[\protect\citeauthoryear{{Irwin} et~al.,}{{Irwin}
  et~al.}{2007}]{2007LeoT}
{Irwin} M.~J.,  et~al., 2007, \mn@doi [\apjl] {10.1086/512183}, \href
  {http://adsabs.harvard.edu/abs/2007ApJ...656L..13I} {656, L13}

\bibitem[\protect\citeauthoryear{{Johnston}, {Bullock}, {Sharma}, {Font},
  {Robertson}  \& {Leitner}}{{Johnston} et~al.}{2008}]{Johnston2008}
{Johnston} K.~V.,  {Bullock} J.~S.,  {Sharma} S.,  {Font} A.,  {Robertson}
  B.~E.,   {Leitner} S.~N.,  2008, \mn@doi [\apj] {10.1086/592228}, \href
  {http://adsabs.harvard.edu/abs/2008ApJ...689..936J} {689, 936}

\bibitem[\protect\citeauthoryear{{Katz} \& {Ricotti}}{{Katz} \&
  {Ricotti}}{2013}]{KatzR2013}
{Katz} H.,  {Ricotti} M.,  2013, \mn@doi [\mnras] {10.1093/mnras/stt676}, \href
  {http://adsabs.harvard.edu/abs/2013MNRAS.432.3250K} {432, 3250}

\bibitem[\protect\citeauthoryear{{Katz} \& {Ricotti}}{{Katz} \&
  {Ricotti}}{2014}]{KatzR2014}
{Katz} H.,  {Ricotti} M.,  2014, \mn@doi [\mnras] {10.1093/mnras/stu1489},
  \href {http://adsabs.harvard.edu/abs/2014MNRAS.444.2377K} {444, 2377}

\bibitem[\protect\citeauthoryear{{Kirby}, {Bullock}, {Boylan-Kolchin},
  {Kaplinghat}  \& {Cohen}}{{Kirby} et~al.}{2014}]{Mdm2014}
{Kirby} E.~N.,  {Bullock} J.~S.,  {Boylan-Kolchin} M.,  {Kaplinghat} M.,
  {Cohen} J.~G.,  2014, \mn@doi [\mnras] {10.1093/mnras/stu025}, \href
  {http://adsabs.harvard.edu/abs/2014MNRAS.439.1015K} {439, 1015}

\bibitem[\protect\citeauthoryear{{Kravtsov}}{{Kravtsov}}{2013}]{Kravtsov2013}
{Kravtsov} A.~V.,  2013, \mn@doi [\apjl] {10.1088/2041-8205/764/2/L31}, \href
  {http://adsabs.harvard.edu/abs/2013ApJ...764L..31K} {764, L31}

\bibitem[\protect\citeauthoryear{{Leaman} et~al.,}{{Leaman}
  et~al.}{2012}]{2012WLM}
{Leaman} R.,  et~al., 2012, \mn@doi [\apj] {10.1088/0004-637X/750/1/33}, \href
  {http://adsabs.harvard.edu/abs/2012ApJ...750...33L} {750, 33}

\bibitem[\protect\citeauthoryear{{McConnachie}}{{McConnachie}}{2012}]{McConnachie2012}
{McConnachie} A.~W.,  2012, \mn@doi [\aj] {10.1088/0004-6256/144/1/4}, \href
  {http://adsabs.harvard.edu/abs/2012AJ....144....4M} {144, 4}

\bibitem[\protect\citeauthoryear{{McLaughlin}}{{McLaughlin}}{1999}]{McLaughlin:99}
{McLaughlin} D.~E.,  1999, \aj, 117, 2398

\bibitem[\protect\citeauthoryear{{Novak}, {Jonsson}, {Primack}, {Cox}  \&
  {Dekel}}{{Novak} et~al.}{2012}]{Novak}
{Novak} G.~S.,  {Jonsson} P.,  {Primack} J.~R.,  {Cox} T.~J.,   {Dekel} A.,
  2012, \mn@doi [\rm{MNRAS}] {10.1111/j.1365-2966.2012.21242.x}, \href
  {http://adsabs.harvard.edu/abs/2012MNRAS.424..635N} {424, 635}

\bibitem[\protect\citeauthoryear{{Okamoto}, {Gao}  \& {Theuns}}{{Okamoto}
  et~al.}{2008}]{OkamotoGT08}
{Okamoto} T.,  {Gao} L.,   {Theuns} T.,  2008, \mn@doi [\mnras]
  {10.1111/j.1365-2966.2008.13830.x}, \href
  {http://adsabs.harvard.edu/abs/2008MNRAS.390..920O} {390, 920}

\bibitem[\protect\citeauthoryear{{Parkinson}, {Cole}  \& {Helly}}{{Parkinson}
  et~al.}{2008}]{Parkinson}
{Parkinson} H.,  {Cole} S.,   {Helly} J.,  2008, \mn@doi [\rm{MNRAS}]
  {10.1111/j.1365-2966.2007.12517.x}, \href
  {http://adsabs.harvard.edu/abs/2008MNRAS.383..557P} {383, 557}

\bibitem[\protect\citeauthoryear{{Planck Collaboration} et~al.,}{{Planck
  Collaboration} et~al.}{2018}]{Planck2018}
{Planck Collaboration} et~al., 2018, preprint, \href
  {http://adsabs.harvard.edu/abs/2018arXiv180706209P} {} (\mn@eprint {arXiv}
  {1807.06209})

\bibitem[\protect\citeauthoryear{{Raimondo}, {Brocato}, {Cantiello}  \&
  {Capaccioli}}{{Raimondo} et~al.}{2005}]{Raimondo2005}
{Raimondo} G.,  {Brocato} E.,  {Cantiello} M.,   {Capaccioli} M.,  2005,
  \mn@doi [\aj] {10.1086/497591}, \href
  {http://adsabs.harvard.edu/abs/2005AJ....130.2625R} {130, 2625}

\bibitem[\protect\citeauthoryear{{Read}, {Iorio}, {Agertz}  \&
  {Fraternali}}{{Read} et~al.}{2017}]{Mdm2017}
{Read} J.~I.,  {Iorio} G.,  {Agertz} O.,   {Fraternali} F.,  2017, \mn@doi
  [\mnras] {10.1093/mnras/stx147}, \href
  {http://adsabs.harvard.edu/abs/2017MNRAS.467.2019R} {467, 2019}

\bibitem[\protect\citeauthoryear{{Ricotti}}{{Ricotti}}{2002}]{Ricotti2002}
{Ricotti} M.,  2002, \mn@doi [\mnras] {10.1046/j.1365-8711.2002.05990.x}, \href
  {http://adsabs.harvard.edu/abs/2002MNRAS.336L..33R} {336, L33}

\bibitem[\protect\citeauthoryear{{Ricotti}}{{Ricotti}}{2009}]{Ricotti2009}
{Ricotti} M.,  2009, \mn@doi [\mnras] {10.1111/j.1745-3933.2008.00586.x}, \href
  {http://adsabs.harvard.edu/abs/2009MNRAS.392L..45R} {392, L45}

\bibitem[\protect\citeauthoryear{{Ricotti} \& {Gnedin}}{{Ricotti} \&
  {Gnedin}}{2005}]{RicottiG2005}
{Ricotti} M.,  {Gnedin} N.~Y.,  2005, \mn@doi [\apj] {10.1086/431415}, \href
  {http://adsabs.harvard.edu/abs/2005ApJ...629..259R} {629, 259}

\bibitem[\protect\citeauthoryear{{Ricotti}, {Gnedin}  \& {Shull}}{{Ricotti}
  et~al.}{2002}]{RicottiGS2002b}
{Ricotti} M.,  {Gnedin} N.~Y.,   {Shull} J.~M.,  2002, \mn@doi [\apj]
  {10.1086/341256}, \href {http://adsabs.harvard.edu/abs/2002ApJ...575...49R}
  {575, 49}

\bibitem[\protect\citeauthoryear{{Ricotti}, {Gnedin}  \& {Shull}}{{Ricotti}
  et~al.}{2008}]{RicottiGS2008}
{Ricotti} M.,  {Gnedin} N.~Y.,   {Shull} J.~M.,  2008, \mn@doi [\apj]
  {10.1086/590901}, \href {http://adsabs.harvard.edu/abs/2008ApJ...685...21R}
  {685, 21}

\bibitem[\protect\citeauthoryear{{Ricotti}, {Parry}  \& {Gnedin}}{{Ricotti}
  et~al.}{2016}]{RicottiPG2016}
{Ricotti} M.,  {Parry} O.~H.,   {Gnedin} N.~Y.,  2016, \mn@doi [\apj]
  {10.3847/0004-637X/831/2/204}, \href
  {http://adsabs.harvard.edu/abs/2016ApJ...831..204R} {831, 204}

\bibitem[\protect\citeauthoryear{{Searle} \& {Zinn}}{{Searle} \&
  {Zinn}}{1978}]{Searle1978}
{Searle} L.,  {Zinn} R.,  1978, \mn@doi [\apj] {10.1086/156499}, \href
  {http://adsabs.harvard.edu/abs/1978ApJ...225..357S} {225, 357}

\bibitem[\protect\citeauthoryear{{Sibbons}, {Ryan}, {Irwin}  \&
  {Napiwotzki}}{{Sibbons} et~al.}{2015}]{2015IC1613}
{Sibbons} L.~F.,  {Ryan} S.~G.,  {Irwin} M.,   {Napiwotzki} R.,  2015, \mn@doi
  [\aap] {10.1051/0004-6361/201423982}, \href
  {http://adsabs.harvard.edu/abs/2015AAp...573A..84S} {573, A84}

\bibitem[\protect\citeauthoryear{{Vansevi{\v c}ius} et~al.,}{{Vansevi{\v c}ius}
  et~al.}{2004}]{2004LeoA}
{Vansevi{\v c}ius} V.,  et~al., 2004, \mn@doi [\apjl] {10.1086/423802}, \href
  {http://adsabs.harvard.edu/abs/2004ApJ...611L..93V} {611, L93}

\bibitem[\protect\citeauthoryear{{White} \& {Frenk}}{{White} \&
  {Frenk}}{1991}]{White1991}
{White} S.~D.~M.,  {Frenk} C.~S.,  1991, \mn@doi [\apj] {10.1086/170483}, \href
  {http://adsabs.harvard.edu/abs/1991ApJ...379...52W} {379, 52}

\bibitem[\protect\citeauthoryear{{Widrow}, {Barber}, {Chequers}  \&
  {Cheng}}{{Widrow} et~al.}{2014}]{2014MNRAS.440.1971W}
{Widrow} L.~M.,  {Barber} J.,  {Chequers} M.~H.,   {Cheng} E.,  2014, \mn@doi
  [\mnras] {10.1093/mnras/stu396}, \href
  {http://adsabs.harvard.edu/abs/2014MNRAS.440.1971W} {440, 1971}

\bibitem[\protect\citeauthoryear{{Wise}, {Turk}, {Norman}  \& {Abel}}{{Wise}
  et~al.}{2012}]{Wise2012}
{Wise} J.~H.,  {Turk} M.~J.,  {Norman} M.~L.,   {Abel} T.,  2012, \mn@doi
  [\apj] {10.1088/0004-637X/745/1/50}, \href
  {http://adsabs.harvard.edu/abs/2012ApJ...745...50W} {745, 50}

\bibitem[\protect\citeauthoryear{{Zolotov}, {Willman}, {Brooks}, {Governato},
  {Brook}, {Hogg}, {Quinn}  \& {Stinson}}{{Zolotov}
  et~al.}{2009}]{2009ApJ...702.1058Z}
{Zolotov} A.,  {Willman} B.,  {Brooks} A.~M.,  {Governato} F.,  {Brook} C.~B.,
  {Hogg} D.~W.,  {Quinn} T.,   {Stinson} G.,  2009, \mn@doi [\apj]
  {10.1088/0004-637X/702/2/1058}, \href
  {http://adsabs.harvard.edu/abs/2009ApJ...702.1058Z} {702, 1058}

\bibitem[\protect\citeauthoryear{{van den Bergh}}{{van den
  Bergh}}{1994}]{vandenBergh1994}
{van den Bergh} S.,  1994, \mn@doi [\aj] {10.1086/116946}, \href
  {http://adsabs.harvard.edu/abs/1994AJ....107.1328V} {107, 1328}

\makeatother
\end{thebibliography}




\appendix
\section{Fits to Simulation Data}\label{app:fit}

\subsection{Radius of Ghostly Stellar Halos}

Here we aim at finding an analytical relationship for the radius of ghostly stellar halos by fitting the simulation data as a function of dark matter halo mass normalized to the Milky Way mass and $\beta$:
\begin{equation}
R^{\rm halo} =R^{\rm halo}_{0}\left(\frac{M_{\rm dm}^{\rm halo}}{M_{\rm mw}}\right)^{\gamma(\beta)}\label{eq:rff},
\end{equation}
where $M_{\rm mw} = 10^{12}$~M$_{\odot}$. We first fit the logarithm of $R^{\rm halo}_{0}$ as a function of $\beta$ with a quadratic polynomial by setting $M_{\rm halo} = M_{\rm mw}$ which yields $R^{\rm halo} = R^{\rm halo}_{0}$:
\begin{equation}
R^{\rm halo}_{0} = 10^{m_0\beta^2+n_0\beta+c_0}.
\end{equation}
For a given $\beta$, $R^{\rm halo} \propto M_{\rm halo}^{\gamma}$, where $\gamma$ depends on $\beta$. We fit $\gamma$ as a function of $\beta$ using quadratic regression. Thus, we obtain the final form of~(\ref{eq:rff}), normalized to $M_{\rm mw}$:
\begin{equation}
R^{\rm halo} = R^{\rm halo}_{0}\left(\frac{M_{\rm dm}^{\rm halo}}{M_{\rm mw}}\right)^{m\beta^2+n\beta+c}\label{eq:rf12}.
\end{equation}
The size of stellar halos depends on the number of mergers and is larger with increasing present day dark matter halo mass. It is interesting to note that for $\beta \sim 0$ the half-light radii of stellar halos vary significantly as a function of the dark matter halo mass while for larger $\beta$ the mass dependence is weaker.

\subsection{Mass of Ghostly Stellar Halos}

We aim to model the mass of ghostly stellar halos as a function of dark matter halo mass normalized to the Milky Way mass and $\beta$:
\begin{equation}
M^{\rm halo}_{*} = M^{\rm halo}_{*,0}(\beta)\left(\frac{M_{\rm dm}^{\rm halo}}{M_{\rm mw}}\right)^{\kappa(\beta)}.
\end{equation}
Going back to Eqs.~(\ref{eq:ms})-(\ref{eq:fs}), we find $M_{*}^{\rm halo} \propto \frac{\epsilon}{M_{\rm mw}^{\beta}}$. Then, we can generalize Eq.~(\ref{eq:ms15}) for any normalization factor as
\begin{equation}
M^{\rm halo}_{*} = \left(\frac{\epsilon^{'}}{\epsilon}\right)\left(\frac{M^{'}}{M_{\rm mw}}\right)^{-\beta} M^{\rm halo}_{*,0}(\beta)\left(\frac{M_{\rm dm}^{\rm halo}}{M_{\rm mw}}\right)^{\kappa(\beta)},
\end{equation}
where $M^{'}$ is the normalization mass and $\epsilon^{'}$ is the corresponding parameter value. 
Our goal is to find $M^{\rm halo}_{*,0}$ and $\kappa(\beta)$. We again set $M_{\rm dm}^{\rm halo} = M_{\rm mw}$ which yields $M_*^{\rm halo} = M_{*,0}^{\rm halo}$:
\begin{equation}
M^{\rm halo}_{*,0} = 10^{p_0 + q_0\beta}.
\end{equation}
Now we fix dark matter halo mass and vary $\beta$. This gives us $M^{\rm halo}_{*} \propto 10^{\lambda\beta}$ where $\lambda$ is a constant that depends on $M_{\rm dm}$. Thus, we find $\kappa(\beta)$:
\begin{equation}
M^{\rm halo}_{*} = M^{\rm halo}_{*,0}\left(\frac{M_{\rm dm}^{\rm halo}}{M_{\rm mw}}\right)^{p + q\beta}\label{eq:ms15}.
\end{equation}
For $\beta = 0$, we see that the stellar halo mass is directly proportional to dark matter mass. As $\beta$ increases, we see a slight non-linearity.  All the fitting parameters for the fiducial run and the two models shown in Figs.~\ref{fig:radius}-\ref{fig:sigma} are listed in Table~\ref{tab:param}.
\begin{table*}
\begin{center}
    \begin{tabular}{ | l | l | l | p{5cm} |}
    \hline
    Parameters & Values   \\ \hline

    $m_0$ & 0.5927\\ \hline
    $n_0$ & -1.8833\\ \hline
    $c_0$ & 2.1003\\ \hline
    $m$ & 0.1474 \\ \hline
    $n$ & -0.4321\\ \hline
    $c$ & 0.6150\\ \hline
    $p_0$  & 8.4932\\ \hline
    $q_0$ & -0.9723\\ \hline
    $p$ & 1.0153\\ \hline
    $q$ & -0.1448\\ \hline
    
    \end{tabular}~~
    \begin{tabular}{ | l | l | l | p{5cm} |}
    \hline
    Parameters & Values   \\ \hline

    $m_0$ & 0.4729\\ \hline
    $n_0$ &  -1.5823\\ \hline
    $c_0$ & 1.9039\\ \hline
    $m$ & 0.1404 \\ \hline
    $n$ & -0.4313\\ \hline
    $c$ & 0.6252\\ \hline
    $p_0$  & 8.4356\\ \hline
    $q_0$ & -0.1906\\ \hline
    $p$ & 1.0253\\ \hline
    $q$ & -0.1444\\ \hline
    
    \end{tabular}~~
    \begin{tabular}{ | l | l | l | p{5cm} |}
    \hline
    Parameters & Values   \\ \hline
    
    $m_0$ & 0.5123\\ \hline
    $n_0$ &  -1.6179\\ \hline
    $c_0$ & 1.8497\\ \hline
    $m$ & 0.1483 \\ \hline
    $n$ & -0.4325\\ \hline
    $c$ & 0.5938\\ \hline
    $p_0$  & 8.5428\\ \hline
    $q_0$ & -0.2227\\ \hline
    $p$ & 1.0130\\ \hline
    $q$ & -0.1686\\ \hline
    
    \end{tabular}
    \caption{Listed are the parameter values for the fitting functions of the effective radii and stellar masses of ghostly halos obtained from the simulation results for $z_{\rm cut} = 6$ and $M_{\rm res} = 10^6$~M$_{\odot}$ (left), $z_{\rm cut} = 6$ and $M_{\rm res} = 5 \times 10^6$~M$_{\odot}$ (middle) and $z_{\rm cut} = 5$ and $M_{\rm res} = 5 \times 10^6$~M$_{\odot}$ (right). The curves showing the fitting functions are shown as solid lines in Figures~\ref{fig:radius}-\ref{fig:sigma}. The stellar mass surface density is derived from the parameter values using Eq.~(\ref{eq:ss}).}\label{tab:param}
\end{center}
\end{table*}
    
\subsection{Mass Surface Density of Ghostly Stellar Halos}

We have fit simulation data to half-light radius and stellar mass models. Therefore, we use these models to derive the stellar mass surface density model using Eq.~(\ref{eq:ss}).
The overall dependence of the stellar mass surface density on $\beta$ and M$_{\rm dm}$ is weaker when compared to the effective radii and stellar mass. For redshift $z \sim 6$, in particular, the dependence on the halo dark matter mass is minimal. 

\subsection{Scatter of Results for Different Random Realizations}

For the purpose of our research, we take the average of 60 simulated realizations. In this section, we explore the spread of the data by looking at the number of mergers and the effective halo radius for a fixed set of parameters but different random realizations. We are particularly interested in the range $\beta \sim -0.5 - 0.5$ and present day dark matter halo mass, $M_{\rm dm}^{\rm halo} \sim 10^8 - 5 \times 10^{10}$~M$_{\odot}$, since our comparison to observations show that these are likely ranges for the star formation efficiency in early galaxies. The number of mergers primarily depends on the mass resolution, $M_{\rm res}$, and $M_{\rm dm}^{\rm halo}$. Lower $M_{\rm res}$ and higher $M_{\rm dm}^{\rm halo}$ allow for more mergers to happen in the simulation. The distribution functions of the number of mergers is Gaussian and $\beta$ has no effect on the shape of the distribution. The scatter of final effective radii is Gaussian when $\beta \sim 0$. However, the data is quite skewed when $\beta$ deviates from $\beta=0$. Tables~\ref{tab:scatter1}-\ref{tab:scatter2} list the fractional scatter of the number of mergers and $r_h^{\rm halo}$ as a function of $\beta$ and M$_{\rm dm}$.
\begin{table*}
\begin{center}
	\centering
    Present Day Dark Matter Halo Mass\\
    \begin{tabular}{lllllllllll}
    \hline
    $\beta$ & $1 \times 10^8 \rm M_{\odot}$ (min, max) & $1 \times 10^9 \rm M_{\odot}$ (min, max) & $1 \times 10^{10} \rm M_{\odot}$ (min, max)\\ 
    & num. of mergers & num. of mergers & num. of mergers\\
\hline
   
   -0.5 & $5.52 \pm 1.86$ $(1, 9)$ & $36.52 \pm 6.89$ $(17, 52)$ & $268.15 \pm 30.62$ $(202, 348)$ \\
   0 & $5.52 \pm 1.86$ $(1, 9)$ & $36.52 \pm 6.89$ $(17, 52)$ & $268.15 \pm 30.62$ $(202, 348)$\\
   0.5 & $5.52 \pm 1.86$ $(1, 9)$ & $36.52 \pm 6.89$ $(17, 52)$ & $268.15 \pm 30.62$ $(202, 348)$\\ 
    \end{tabular}
    \begin{tabular}{lllllllllll}
    \hline
    $\beta$ & $5 \times 10^8 \rm M_{\odot}$ (min, max) & $5 \times 10^9 \rm M_{\odot}$ (min, max) & $5 \times 10^{10} \rm M_{\odot}$ (min, max)\\
    & num. of mergers & num. of mergers & num. of mergers\\
\hline
   
   -0.5 & $5.33 \pm 1.60$ $(1, 8)$ & $35.67 \pm 5.86$ $(22, 50)$ & $270.07 \pm 25.57$ $(216, 334)$ \\
   0 & $5.33 \pm 1.60$ $(1, 8)$ & $35.67 \pm 5.86$ $(22, 50)$ & $270.07 \pm 25.57$ $(216, 334)$\\
   0.5 & $5.33 \pm 1.60$ $(1, 8)$ & $35.67 \pm 5.86$ $(22, 50)$ & $270.07 \pm 25.57$ $(216, 334)$\\
    \end{tabular}
    \caption{The average number of halo mergers from the 60 realizations of merger history with errors, minimum, and maximum at $z = 6$, $M_{res} = 10^6 M_{\odot}$ (top) and $M_{res} = 5 \times 10^6 M_{\odot}$ (bottom). The number of mergers is independent of $\beta$. Instead, it primarily depends on present day dark matter halo mass. The bottom table starts from $5 \times 10^8 \rm M_{\odot}$ due to higher mass resolution.}\label{tab:scatter1}
\end{center}    
\end{table*}

\begin{table*}
\begin{center}
	\centering
    Present Day Dark Matter Halo Mass\\
    \begin{tabular}{lllllllllll}
    \hline
    $\beta$ & $1 \times 10^8 \rm M_{\odot}$ (min, max) & $1 \times 10^9 \rm M_{\odot}$ (min, max) & $1 \times 10^{10} \rm M_{\odot}$ (min, max)\\
    & $\rm [kpc]$ & $\rm [kpc]$ & $\rm [kpc]$\\

\hline
   
   -0.5 & $0.50 \pm 24\%$ $(0.17, 0.74)$ & $3.45 \pm 17\%$ $(1.74, 4.75)$ & $24.99 \pm 12\%$ $(18.28, 33.82)$ \\
   0 & $0.42 \pm 26\%$ $(0.17, 0.65)$ & $1.74 \pm 32\%$ $(0.65, 2.72)$ & $7.10 \pm 34\%$ $(3.15, 15.82)$\\
   0.5 & $0.33 \pm 27\%$ $(0.17, 0.55)$ & $0.90 \pm 38\%$ $(0.42, 1.62)$ & $2.50 \pm 41\%$ $(1.06, 6.52)$\\ 
    \end{tabular}
    \begin{tabular}{lllllllllll}
    \hline
    $\beta$ & $5 \times 10^8 \rm M_{\odot}$ (min, max) & $5 \times 10^9 \rm M_{\odot}$ (min, max) & $5 \times 10^{10} \rm M_{\odot}$ (min, max)\\
    & $\rm [kpc]$ & $\rm [kpc]$ & $\rm [kpc]$\\
\hline
   
   -0.5 & $0.80 \pm 23\%$ $(0.31, 1.16)$ & $5.64 \pm 16\%$ $(3.62, 7.44)$ & $42.64 \pm 10\%$ $(32.10, 52.64)$ \\
   0 & $0.69 \pm 25\%$ $(0.31, 1.11)$ & $3.03 \pm 31\%$ $(1.20, 4.80)$ & $12.86 \pm 30\%$ $(5.26, 21.25)$\\
   0.5 & $0.55 \pm 27\%$ $(0.31, 1.03)$ & $1.62 \pm 39\%$ $(0.69, 3.04)$ & $4.35 \pm 40\%$ $(1.50, 9.45)$\\
    \end{tabular}
    \caption{Same as Table~\ref{tab:scatter1}, but for the spread of radii. Here the standard deviation are presented as relative error in per cents. Unlike the number of mergers, the scatter of the radii is not Gaussian (skewed) when $\beta$ significantly deviates from zero.}\label{tab:scatter2}
\end{center}    
\end{table*}

\section{CMD Diagrams for Isolated Dwarfs}\label{app:cmd}

\begin{figure*}
\centering
\includegraphics[width=0.48\textwidth]{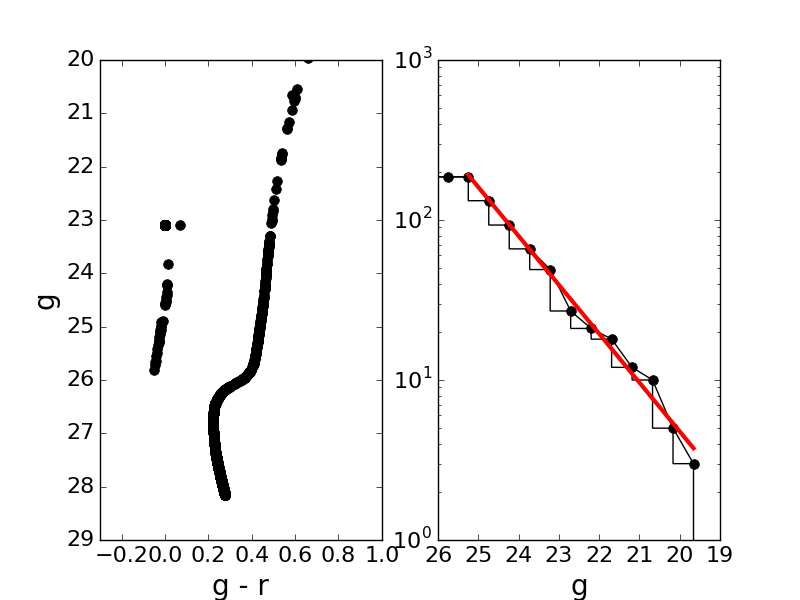}
\includegraphics[width=0.48\textwidth]{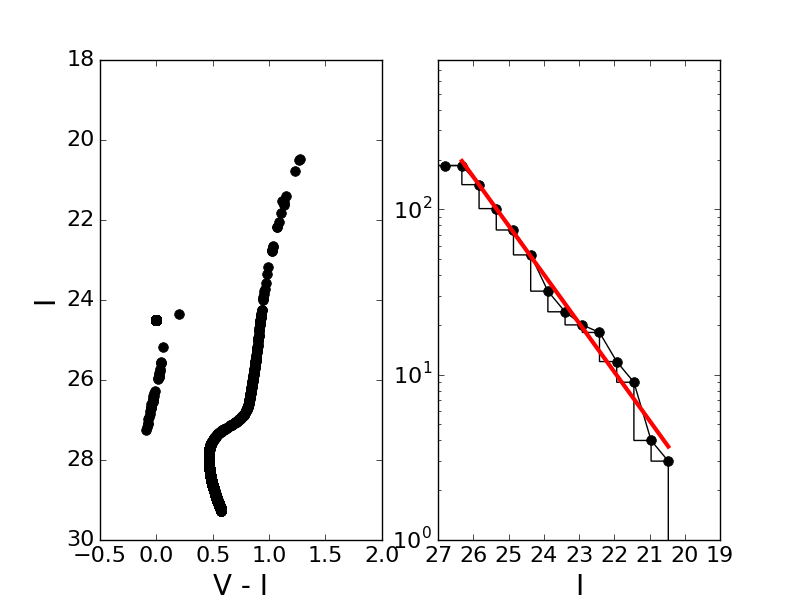}\\
\includegraphics[width=0.48\textwidth]{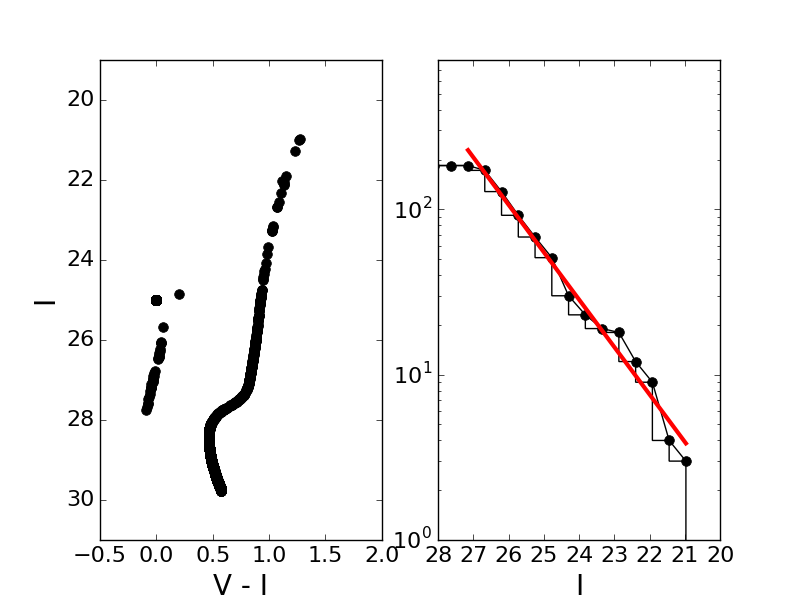}
\includegraphics[width=0.48\textwidth]{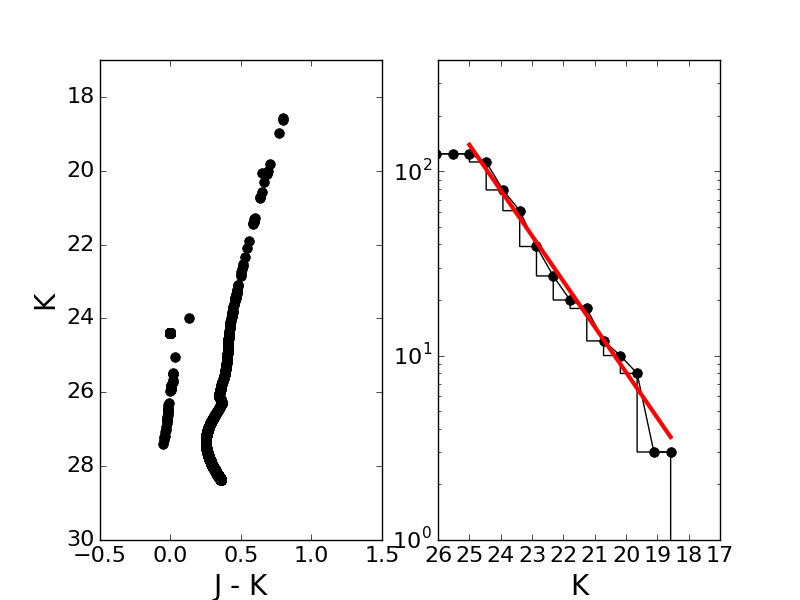}\\
\includegraphics[width=0.48\textwidth]{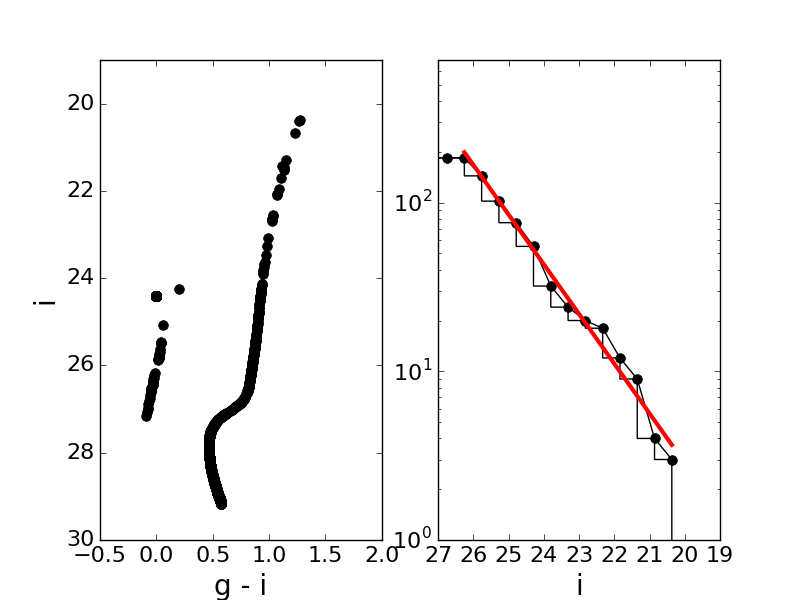}
\includegraphics[width=0.48\textwidth]{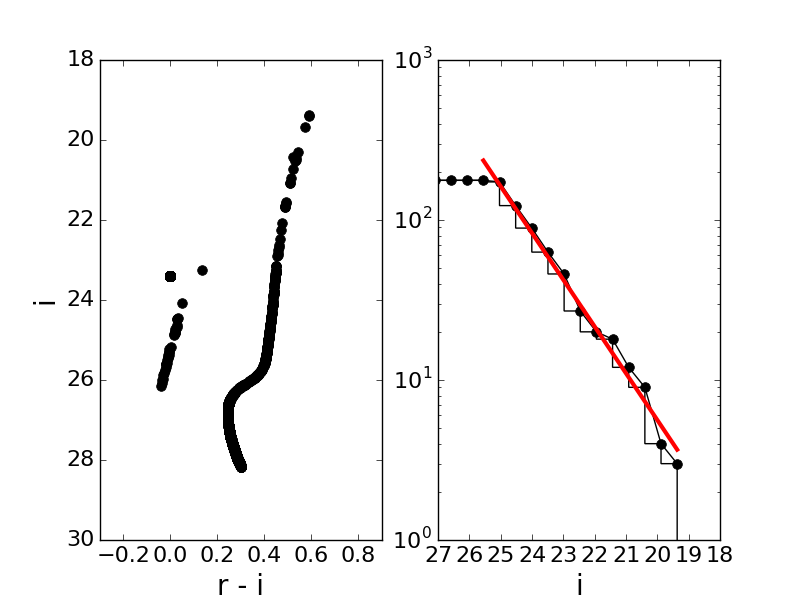}
\caption{ Synthetic color-magnitude diagrams (CMD) for the six dwarfs considered in this study. The left panels in each of the six figures show the CMD for the same filters as in the data from the literature review. The right panels show the cumulative number of stars along the red-giant branch as a function of the faintest magnitude cut in the data. From left to right and top to bottom, the CMD are for LeoT (stellar population with $g < 24.3$ and $0.2 < g - r$  2); LeoA ($I < 23$ and $0.8 < V - I$); WLM ($V < 23.5$ and $0.6 < V - I$ 4); IC10 ($I < 23.5$ and $0.6 < V - I$); NGC6822 ($i < 24$ and $0.2 < r - i$).}
\label{fig:cmd}
\end{figure*}

The purpose of this section is to estimate the stellar mass of ghostly halos of the six nearby dwarf galaxies in Table~\ref{tab:gal1}. To do so we use synthetic CMD models to produce simple stellar populations (single-age, single-metallicity) with a metal poor ($Z = 0.0003$) and an old stellar population (age of 13 billion years) \citep{Brocato2000,Brocato1999,Cantiello2003,Raimondo2005}. We set limits, $m_{\rm lim}$, based on CMD's of RGB/AGB stars in our chosen dwarf galaxies shown in Table~\ref{tab:cmd} and use cumulative sum to find linear regression shown in Fig.~\ref{fig:cmd}. We rely on synthetic CMD models to select RGB/AGB stars while we use $m_{\rm lim}$ to limit the number of stars. Finally, we calculate the stellar mass using the following equation: 
\begin{align}
N^{\rm RGB/AGB} = 10^{a + bm_{\rm lim}}\left(\frac{M_*}{2.8 \times 10^4~{\rm M}_\odot}\right),
\end{align}
where $N^{\rm RGB/AGB}$ represents the number of stars from galaxies and $2.8 \times 10^4$~M$_\odot$ is the estimated total stellar mass from our synthetic CMD models.

\begin{table*}
\begin{center}
    \begin{tabular}{lllllllllll}
    \hline
    Galaxy & $r_{cut}$ & CMD cut & $m_{\rm lim}$ & CMD param. & CMD\\ 
    & [kpc] & & & $(a, b)$ & $10^{a + bm_{\rm lim}}$\\ 
\hline
\\   
   Leo T & 0.24 & $g < 24.3$, $0.4 < g - r$ & 24.3 & (-5.43, 0.31) & 97.95 \\
   Leo A & 1.28 & $I < 23$, $0.8 < V - I$ & 23 & (-5.47, 0.29) & 20.48 \\
   WLM & 3.58 & $I < 23.5$, $0.8 < V - I$ & 23.5 & (-5.43, 0.29) & 20.44 \\
   IC1613 & 4.5 & $K < 18.28$, $0.4 < J - K$ & 18.28 & (-4.02, 0.25) & 3.06 \\
   IC10 & 3.27 & $i < 22.5$, $0.8 < g - i$ & 22.5 & (-5.43, 0.29) & 15.56 \\
   NGC6822 & 3.28 & $i< 24$, $0.4 < r - i$ & 24 & (-5.09, 0.29) & 82.63 \\
   
    \end{tabular}
    \caption{The radial cut for the halo $r_{\rm cut}$, the color cuts and the limiting apparent magnitude $m_{\rm lim}$ used in the references previously noted in Figure~\ref{fig:profs}. The radius $r_{\rm cut}$ is the estimated cutoff point between galactic and halo stars and $m_{\rm lim}$ is the limiting apparent magnitude due to foregrounds and sensitivity limitations. CMD color cuts are also used to select RGB/AGB stars in the synthetic CMD models and limit foreground contamination. The parameters to determine the number of star in the red-giant branch are calculated using synthetic CMD data shown in Fig.~\ref{fig:cmd}(right panel) with color cuts, and $m_{\rm lim}$ reported in this table.}\label{tab:cmd}
    \end{center}    
\end{table*}


\bsp	
\label{lastpage}
\end{document}